\documentclass[12pt,a4paper]{article}
\usepackage{mathrsfs}
\usepackage{epsfig}
\usepackage{slashed}
\usepackage{amssymb}
\usepackage{amsmath}
\usepackage{amsthm}
\begin{document}

\title{Probe Anomalous $\rm tq\gamma$ couplings through Single Top Photoproduction at the LHC}
\author{Hao Sun\footnote{haosun@mail.ustc.edu.cn \hspace{0.2cm} haosun@dlut.edu.cn} \\
{\small Institute of Theoretical Physics, School of Physics $\&$ Optoelectronic Technology,} \\
{\small Dalian University of Technology, Dalian 116024, P.R.China}
}
\date{}
\maketitle

\vspace{-0.5cm}
\begin{abstract}
In this work we study the constraints on the anomalous $\rm tq\gamma$ (q=u, c) couplings
by photon-produced leading single top production
and single top jet associated production through the main
reaction $\rm pp\rightarrow p\gamma p\rightarrow pt\rightarrow pW(\rightarrow\ell \nu_\ell) b+X$
and $\rm pp\rightarrow p\gamma p\rightarrow ptj\rightarrow pW(\rightarrow\ell \nu_\ell) bj+X$
assuming a typical LHC multipurpose forward detectors in a model independent effective
lagrangian approach. Our results show that:
for the typical detector acceptance $0.0015<\xi_1<0.5$, $0.1<\xi_2<0.5$ and $0.0015<\xi_3<0.15$
with a luminosity of 2 $\rm {fb}^{-1}$, the lower bounds of $\rm \kappa_{tq\gamma}$
through leading single top channel (single top jet channel) are
0.0130 (0.0156), 0.0218 (0.0206) and 0.0133 (0.01655), respectively,
correspond to $\rm Br(t\rightarrow q\gamma)\sim 3\times 10^{-5}$.
With a luminosity of 200 $\rm fb^{-1}$, the lower bounds of $\rm \kappa_{tq\gamma}$ are
0.0041 (0.0048), 0.0069 (0.0064) and 0.0042 (0.0051), respectively,
correspond to $\rm Br(t\rightarrow q\gamma)\sim 4\times 10^{-6}$.
We conclude that both channels can be used to detect such anomalous $\rm tq\gamma$ couplings
and the detection sensitivity on $\rm \kappa_{tq\gamma}$ is obtained.

\vspace{-0.5cm} \vspace{2.0cm} \noindent
 {\bf Keywords}: Anomalous Coupling, Forward Detector, Large Hadron Collider  \\
 {\bf PACS numbers}: 12.60.-i, 14.65.Ha
\end{abstract}

\newpage
\section{Introduction}
The top quark is the heaviest known elementary particle which makes it
an excellent candidate for new physics searches.
One possible manifestation of new interaction in the top quark sector
is to alter its couplings to the gauge bosons.
Such anomalous couplings would modify top production and decay at colliders.
The most widely studied cases are the $\rm t\bar{t}V$, with $\rm V=\gamma$,
Z, g, and $\rm t\bar{b}W$ three-point functions.
In addition, the flavor change neutral current (FCNC) interactions tqV,
with q=u, c, will also offer an ideal place to search for new physics.
They are very small in the Standard Model (SM).
For instance, while radiative B-menson decays have branching ratios of
order $\rm Br(b\rightarrow s\gamma) \sim 10^{-4}$, typical FCNC top quark decays,
such as $\rm t\rightarrow cZ$, $\rm t\rightarrow c\gamma$ and $\rm t\rightarrow cg$,
are highly suppressed by GIM mechanism with SM branching ratios of order
at most $10^{-14}$, $10^{-13}$ and $10^{-12}$\cite{tqv_SMTHDM1,tqv_SMTHDM2}, respectively,
which in practice are impossible to be measured. In this instance any positive
observation of these transitions would be signal presence of new physics.
Actually, $\rm t\rightarrow cV$ have been studied in various new
physics models beyond the SM
\cite{
tqv_SMTHDM1,tqv_SMTHDM2,
tqv_THDM1,tqv_THDM2,tqv_THDM3,
tqv_TC,tqv_quarksinglet,
tqv_MSSM1,tqv_MSSM2,tqv_MSSM3,tqv_MSSM4,tqv_MSSM5,tqv_MSSM6,tqv_MSSM7,tqv_MSSM8,
tqv_warpED,tqv_LRSUSY}.
There they often predict much larger FCNC top quark decay interactions
which can be explored in future collider experiments.

In addition to the direct top quark decays, production of top quarks by
FCNC interactions can also be used to probe such vertices. Studies have been
presented at linear colliders
\cite{tqv_ILC1,tqv_ILC2,tqv_ILC3,tqv_ILC4,tqv_ILC5,tqv_ILC6,tqv_hadronee},
lepton-hadron colliders\cite{tqv_ep1, tqv_ep2, tqv_ep3, tqv_ep4},
as well as hadron colliders
\cite{tqv_hadronee,tqv_pp1,tqv_pp2,tqv_pp3,tqv_pp4,tqv_pp5,tqv_pp6,tqv_pp7,tqv_pp8,tqv_pp9,tqv_pp10,
tqv_ppNLO1,tqv_ppNLO2}, see also reference there in.
In this paper, we study the $\rm tq\gamma$ anomalous couplings through the
leading single top photoproduction and single top jet associated
photoproduction via the main reaction
$\rm pp\rightarrow p\gamma p\rightarrow pt\rightarrow pW(\rightarrow\ell\nu_\ell)b+X$
and $\rm pp\rightarrow p\gamma p\rightarrow ptj\rightarrow pW(\rightarrow\ell\nu_\ell)bj+X$
assuming a typical LHC multipurpose forward detectors in a model independent
effective lagrangian approach. Feynman diagrams for these processes present
with anomalous $\rm tq\gamma$ couplings arise from the initial photon. Similar studies
was presented in Ref.\cite{tqv_ep1} and tried to study $\rm tq\gamma$ coupling
through $\rm \gamma b \rightarrow Wb$ at CLIC+LHC ep colliders while recently
moved to the photon-proton ($\rm \gamma p$) collision in Ref.\cite{Anomaloustqr}.
In addition, feasibility studies of anomalous $\rm \kappa_{tq\gamma}$ via single top
photoproduction at the LHC have also been carried out
in Ref.\cite{rbWtVtb1,rbWtVtb2,AnomalousWtb}.
Typically, our study will also include the single top jet associated production channel.

Photon-induced processes have been measured by CDF collaboration, i.e.,
the exclusive lepton pairs production\cite{ppllpp1,ppllpp2},
photon photon production\cite{pprrpp}, dijet production\cite{ppjjpp}
and charmonium ($\rm J/\psi$) meson photoproduction\cite{ppJPHIpp}, etc,
through photon-photon ($\gamma\gamma$) or $\rm \gamma p$ interactions.
Studies of these leptons, photon and heavy particle productions might be
possible and open new field of studying $\gamma\gamma$ and $\rm \gamma p$
collisions with very high energy but very low backgrounds. Indeed,
ATLAS and CMS collaborations have programs of forward physics
with extra updated detectors located in a region nearly 100m-400m
close to the interaction point\cite{FDs1,FDs2}. Technical details
of the ATLAS Forward Physics (AFP) projects can be found, for example,
in Ref.\cite{AFP}. A brief review of experimental prospects for studying
photon induced interactions are summaries in Ref.\cite{HEPhotonIntatLHC}.
By using forward detector equipment one can eliminate many serious backgrounds
and this is one of the advantages of studying the photoproduction processes.
We summarize some phenomenological studies on photon produced processes here:
standard model productions\cite{rbWtVtb1,rbWtVtb2,SMWH},
supersymmetry\cite{SUSYprrp1,SUSYprrp2},
extra dimensions\cite{EDpllp1,EDprrp2,EDrqrq3},
unparticle physics\cite{unparticle}, top triangle moose model\cite{TTMrbtp},
gauge boson self-interactions
\cite{anoWWr1,anoWWr2,anoVVV,anoWWrr,anoWWr3,anoZZZ,anoZZrr,anoZZrZrr,anoWWrrZZrrrp,anoVVVV},
neutrino electromagnetic properties\cite{electromagnetic1,electromagnetic2,electromagnetic3},
the top quark physics\cite{Anomaloustqr,rbWtVtb1,rbWtVtb2,AnomalousWtb}
and triplet Higgs production\cite{TripletH}, etc.

The possibility of adding forward proton detectors to both the ATLAS
and CMS experiments has received quite some attention since the
possibility of forward proton tagging would provide a very clean
environment for new physics searches. Our paper is organized as follow:
we build the calculation framework in Section 2 include a brief
introduction to the anomalous $\rm tq\gamma$ couplings, Equivalent Photon
Approximation implementation, general photoproduction cross section.
Section 3 is arranged to present the selected
processes and numerical results as well as the signal and background analysis.
In Section 4 we present the bounds on anomalous $\rm tq\gamma$ couplings
at the future LHC. Finally we summarize our conclusions in the last section.

\section{Calculation Framework}

The effective Lagrangian involving anomalous $\rm tq\gamma$ (q=u, c) couplings
can be written as
\begin{eqnarray}\label{lagrangian}
\rm {\cal L} = i e e_t \bar{t} \frac{\sigma_{\mu\nu}q^\nu}{\Lambda} \kappa_{tu\gamma} u A^\mu
 +  i e e_t \bar{t} \frac{\sigma_{\mu\nu}q^\nu}{\Lambda} \kappa_{tc\gamma} c A^\mu + h.c.,
\end{eqnarray}
where $\Lambda$ is an effective scale which we set equal to the top quark mass
$\rm m_t$ by convention. Usually the value of $\Lambda$ should be at the TeV scale.
For the other choice of $\Lambda$ the results can be rescaled by $\rm (\frac{m_t}{\Lambda})^2$.
$\rm e$, $\rm e_t$ are the electric charge of the electron and the top quark, respectively.
$\sigma_{\mu\nu}$ is defined as $(\gamma_{\mu}\gamma_{\nu}-\gamma_\nu \gamma_\mu)/2$
with $\gamma_\mu$ the Dirac matrices. $\rm q^\nu$ is the photon 4-vector momentum.
$\rm \kappa_{tu\gamma}$ and $\rm \kappa_{tc\gamma}$ are real and positive anomalous FCNC couplings.

As the SM predictions for $\rm \Gamma(t\rightarrow q\gamma)$ are exceedingly small,
we need only consider $\rm t\rightarrow q\gamma$ decays
mediated by the anomalous $\rm tq\gamma$ interactions,
which can be considered at the next-to-leading order (NLO)\cite{tqr_NLO}
and resulted for the final decay widths $\rm \Gamma(t\rightarrow q\gamma)$:
\begin{eqnarray}
\rm \Gamma(t\rightarrow q\gamma)=\Gamma_{0}(t\rightarrow q\gamma)
\{ 1+\frac{\alpha_s}{\pi} \left[-3 ln \left(\frac{\mu^2}{m_t^2}\right) -2\pi^2+8 \right] \}
\end{eqnarray}
with the leading order (LO) decay width obtained from Eq.(\ref{lagrangian}) as
$\rm \Gamma_0(t\rightarrow q\gamma)=\frac{2}{9} \alpha_{ew} m^3_t \frac{\kappa_{tq\gamma}^2}{\Lambda^2}$
with $\rm \alpha_{ew}=\frac{1}{137}$.
It is convenient to relate the branching ratios $\rm Br(t\rightarrow q\gamma)$
to the FCNC partial widths of the top-quark as
\begin{eqnarray}\label{branchtqvr}
\rm Br(t\rightarrow q\gamma) = \frac{\Gamma(t\rightarrow q\gamma)}{\Gamma(t\rightarrow W^+b)}.
\end{eqnarray}
The decay width of the dominant top-quark decay mode $\rm t\rightarrow Wb$ at the LO and the NLO
could be found in Ref.\cite{twb_NLO}, and is given below
\begin{eqnarray}\label{GammaT} \nonumber
\rm \Gamma(t\rightarrow bW)=\Gamma_0(t\rightarrow bW) \{ 1+\frac{2\alpha_s}{3\pi}
 [2(\frac{(1-\beta^2_W)(2 \beta^2_W-1)(\beta^2_W-2)}{\beta^4_W(3-2\beta^2_W)}) \mbox{ln}(1-\beta^2_W)  \\
\rm - \frac{9-4\beta^2_W}{3-2\beta^2_W} \mbox{ln}\beta^2_W +2Li_2(\beta^2_W)
 -2Li_2(1-\beta^2_W) - \frac{6\beta^4_W-3 \beta^2_W -8}{2 \beta^2_W (3-2\beta^2_W)} -\pi^2 ] \},
\end{eqnarray}
where $\rm \Gamma_0(t\rightarrow bW)=\frac{G_F m_t^3}{8\sqrt{2}\pi}|V_{tb}|^2 \beta^4_W (3-2\beta^2_W)$
is the LO decay width and $\rm \beta_W=(1-m_W^2/m_t^2)^{\frac{1}{2}}$
is the velocity of the W-boson in the top-quark rest frame.

Present constraints on the FCNC $\rm tq\gamma$ couplings come from
the following experimental bounds: The CDF collaboration\cite{tqr_CDF}
has performed a direct search for FCNC top decays and has placed the
$95\%$ confidence level (C.L.) limits on the branching fractions
$\rm Br(t\rightarrow q\gamma)<3.2\%$ (q=u, c), which gives the constraint
of $\rm \kappa_{tq\gamma}\leq 0.26$. The ZEUS collaboration\cite{tqr_ZEUS}
provide at $95\%$ C.L. the effective FCNC coupling $\rm \kappa_{tu\gamma} < 0.174$
with the assumption of $\rm m_t=175~{\rm GeV}$. The current limits from H1
collaboration are $\rm \kappa_{tq\gamma} < 0.305$\cite{tqr_H1}.
These constraints will be improved significantly by the large top quark
sample to be available at the LHC. In particular, both the
ATLAS\cite{anomalous_ATLAS} and CMS\cite{anomalous_CMS} collaborations
have presented their sensitivity to these rare top quark decays induced
by the anomalous FCNC interactions\cite{anomalous_LHC}.

In addition, both the ATLAS and CMS collaborations are considering the
possibility of adding forward proton detectors in experiments.
Different from usual proton-proton (pp) Deep Inelastic Scattering (DIS),
incoming protons dissociate into partons, jets will be made from the proton
remnants which create some ambiguities and make the new physics signal
detection suffer from incredible backgrounds, $\gamma\gamma$ and $\rm \gamma p$
collisions can provide more clean environment. In this case,
the quasi-real photons emitted at very low virtuality from protons,
leave the radiating proton intact, thus providing an extra experimental handle
(forward proton tagging) to help reduce the backgrounds.
Study the sensitivity of the anomalous
FCNC interactions on the $\gamma \gamma$ or $\rm \gamma p$ collisions will give
complementary information for normal pp collisions. Deflected protons and
their energy loss will be detected by the forward detectors with a very large
pseudorapidity. Photons emitted with small angles by the protons show a
spectrum of virtuality $\rm Q^2$ and the energy $\rm E_\gamma$. This is described by
the Equivalent Photon Approximation (EPA)\cite{EPA} which differs from the
point-like electron (positron) case by taking care of the electromagnetic
form factors in the equivalent $\gamma$ spectrum and effective $\gamma$
luminosity:
\begin{equation}
\rm \frac{dN_\gamma}{dE_\gamma dQ^2}=\frac{\alpha}{\pi}\frac{1}{E_\gamma Q^2}[(1-\frac{E_\gamma}{E})(1-\frac{Q^2_{min}}{Q^2})F_E
 + \frac{E^2_\gamma}{2 E^2}F_M]
\end{equation}
with
\begin{eqnarray} \nonumber
\rm Q^2_{min}=\frac{M^2_p E^2_\gamma}{E(E-E_\gamma)}, ~~~~ F_E= \frac{4 M^2_p G^2_E + Q^2 G^2_M}{4 M^2_p +Q^2}, \\
\rm G^2_E=\frac{G^2_M}{\mu^2_p}=(1+\frac{Q^2}{Q^2_0})^{-4}, ~~~~F_M=G^2_M, ~~~~Q^2_0=0.71 GeV^2 ,
\end{eqnarray}
where $\alpha$ is the fine-structure constant, E is the energy of the incoming
proton beam which is related to the quasi-real photon energy by $\rm E_\gamma=\xi E$
and $\rm M_p$ is the mass of the proton. $\rm \mu^2_p$ = 7.78 is the magnetic moment of
the proton. $\rm F_E$ and $\rm F_M$ are functions of the electric and magnetic form factors.
The intact protons with some momentum fraction loss is described by the formula
$\rm \xi = (|p| - |p'|)/|p| $, which is defined as the forward detector acceptances.

We denote the photoproduction processes as
\begin{eqnarray}
\rm pp \rightarrow  p\gamma p \rightarrow p + \gamma + q/\bar{q}/g  \rightarrow p + i + j + k + ... + X
\end{eqnarray}
with q=u, d, c, s, b and i, j, k, ... the final state particles.
The hadronic cross section at the LHC can be converted by integrating
$\rm \gamma + q/\bar{q}/g \rightarrow i + j + k +...$ over the photon ($\rm dN(x,Q^2)$),
gluon and quark ($\rm G_{g,q/p}(x_2,\mu_f)$) spectra:
\begin{eqnarray} \nonumber
\rm \sigma=\int^{\sqrt{\xi_{max}}}_{\frac{M_{inv}}{\sqrt{s}}} 2z dz \int^{\xi_{max}}_{Max(z^2,\xi_{min})}
\frac{dx_1}{x_1} \int^{Q^2_{max}}_{Q^2_{min}} \frac{dN_\gamma(x_1)}{dx_1dQ^2} G_{g,q/p}(\frac{z^2}{x_1}, \mu_f) \\
\rm \cdot \int \frac{1}{avgfac} \frac{|{\cal M}_n ( \hat s =z^2 s
)|^2}{2 \hat s (2 \pi)^{3n-4}} d\Phi_n ,
\end{eqnarray}
where $\rm x_1$ is the ratio between scattered quasi-real photons and incoming proton energy
$\rm x_1 = E_\gamma/E$ and $\rm \xi_{min} (\xi_{max})$ are its lower (upper) limits.
$\rm x_2$ is the momentum fraction of the proton momentum carried by the gluon (quark).
The quantity $\rm \hat s = z^2 s$ is the effective center-of-mass system (c.m.s.)
energy with $\rm z^2=x_1 x_2$. $\rm M_{inv}$
is the total mass of the related final states. $\rm 2z/x_1$ is the Jacobian determinant
when transform the differentials from $\rm dx_1dx_2$ into $\rm dx_1dz$. $\rm G_{g,q/p}(x,\mu_f)$
represent the gluon (quark) parton density functions, $\rm \mu_f$ is the factorization scale.
$\rm \frac{1}{avgfac}$ is the times of spin-average factor, color-average factor
and identical particle factor. $\rm |{\cal M}_n|^2$ presents the squared n-particle matrix element
and divided by the flux factor $\rm [2 \hat s (2 \pi)^{3n-4}]$.
$\rm d\Phi_n$ and $\rm \Phi_n$ are the n-body phase space differential and its integral
depending on $\rm \hat s$ and particle masses.

\section{The Processes and Numerical Results}

We implement the anomalous interaction vertices deduced from the Lagrangian
(see in Eq.(\ref{lagrangian})) into FeynArts and use FeynArts,
FormCalc and LoopTools (FFL) packages\cite{FeynArts,FormCalc,LoopTools}
to create the amplitudes and perform the numerical calculation for both the
signal and background.
We adopt CT10\cite{CT10} PDF for the parton distributions for collider physics
and BASES\cite{BASES} to do the phase space integration while Kaleu\cite{Kaleu}
to cross check. We take the input parameters as $\rm M_p=0.938272046\ GeV$,
$\rm \alpha_{ew}(m^2_Z)^{-1}|_{\overline{MS}}$=127.918,
$\rm m_Z=91.1876\ GeV$, $\rm m_W=80.385\ GeV$\cite{2012PDG}
and we have $\rm \sin^2\theta_W=1-(m_W/m_Z)^2=0.222897$.
For the strong coupling constant $\rm \alpha_s$, we take $\rm \alpha_s$ = 0.118.
We set the factorization scale to be $\rm \mu_f = \mu_0 = m_t/2$.
Throughout this paper, we set the quark masses as $\rm m_u=m_d=m_c=m_s=m_b=0\ GeV$.
$\rm m_e=0.510998910\ MeV$, $\rm m_\mu=105.658367\ MeV$.
The top quark pole mass is set to be $\rm m_t=173.5\ GeV$.
By taking $\rm \alpha_{ew}(m^2_Z)^{-1}|_{\overline{MS}}=127.918$
and $\rm \alpha_s(m^2_t)=0.1079$, we obtain $\rm \Gamma_t=1.41595\ GeV$ from Eq.(\ref{GammaT}).
The colliding energy in the pp c.m.s.
is assumed to be $\rm \sqrt{s}=14\ TeV$ at future LHC with its luminosity
taken to be a running parameter.
Based on the forward proton detectors to be installed by the CMS-TOTEM and
the ATLAS collaborations we choose the detected acceptances to be\cite{anoWWr1,anoWWr2,xi123}
\begin{itemize}
 \item $\rm CMS-TOTEM$ forward detectors with $0.0015<\xi_1<0.5$
 \item $\rm CMS-TOTEM$ forward detectors with $0.1<\xi_2<0.5$
 \item $\rm AFP-ATLAS$ forward detectors with $0.0015<\xi_3<0.15$
\end{itemize}
which we simply refer to $\xi_1$, $\xi_2$ and $\xi_3$, respectively.

\subsection{Direct Single Top Photoproduction}

\begin{figure}[hbtp]
\vspace{-5cm}
\hspace*{-3cm}
\includegraphics[scale=0.9]{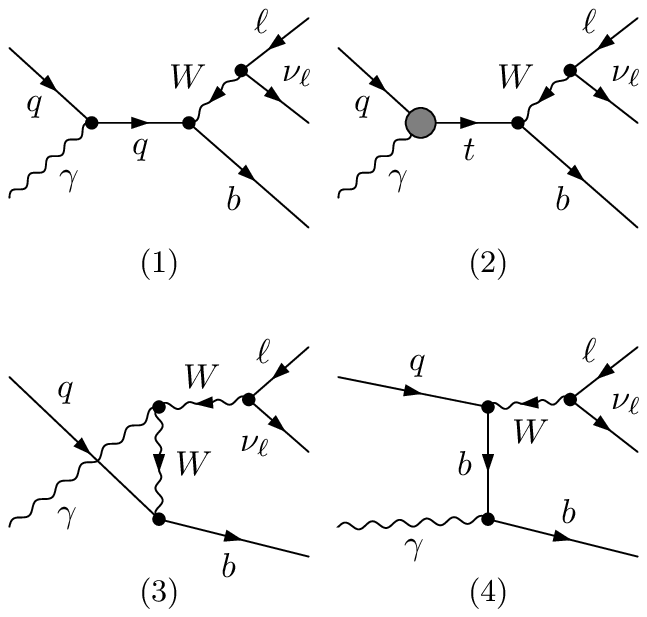}
\vspace{-17cm}
\caption{\label{figrqwb} Partonic Feynman diagrams for
$\rm \gamma q \rightarrow W(\rightarrow \ell\nu_\ell) + b$ with q=u, c.
Black blobs represent the anomalous $\rm tq\gamma$ couplings parameterized
by Eq.(\ref{lagrangian}).}
\end{figure}

The first single top photoproduction with anomalous $\rm tq\gamma$ interactions
we consider is the direct leading single top production via the process
\begin{eqnarray}
\rm pp\rightarrow p\gamma p\rightarrow p t\rightarrow pW(\rightarrow \ell\nu_\ell)b +X
\end{eqnarray}
where q=u, $\rm \bar{u}$, c, $\rm \bar{c}$.
The Feynman diagrams for the subprocess
$\rm \gamma q\rightarrow t\rightarrow W(\rightarrow \ell\nu_\ell)b$ is presented
in Fig.\ref{figrqwb}(2) corresponds to the signal and Fig.\ref{figrqwb}(1,3,4)
correspond to its irreducible background. The black blobs in these figures
represent the anomalous $\rm tq\gamma$ couplings parameterized by Eq.(\ref{lagrangian})
and the anomalous FCNC couplings $\rm \kappa_{tu\gamma}$ ($\rm \kappa_{tc\gamma}$).
Full effects of the top quark leptonic decay modes
($\rm t\rightarrow Wb \rightarrow \ell \nu_{\ell} b$, with $\rm \ell=e$, $\mu$)
are taken into account ($\tau$ leptons are ignored).

In this case, the studied topology is simply one of a tagged b-jet, one isolated,
either positive or negative, lepton $\ell^\pm$, and a missing transverse momentum
from the undetected neutrino. In addition to the irreducible
background from Fig.\ref{figrqwb}(1,3,4), the main background comes from associated
production of W boson and the light jets with jet faking a b-jet. Though jet charge
can be a possibility for labeling jets, it is not well measured experimentally,
we can not use charge to separate them. In our analysis, we assume a b-jet tagging
efficiency of $\rm \epsilon_b=60\%$ and a corresponding mistagging rate of $\rm \epsilon_{light}=1\%$
for light jets (u, d, s quark or gluon) and $\rm \epsilon_c=10\%$ for a c-jet,
consistent with typical values assumed by the LHC experiments\cite{misjets}.

For the direct leading single top production, we impose a cut of pseudorapidity
$|\eta| < 2.5$ for the final state particles since central detectors of the
ATLAS and CMS have a pseudorapidity coverage 2.5. The general acceptance
cuts for both the signal and background events are:
\begin{eqnarray}\label{basiccuts} \nonumber
&&\rm p_T^{jet} \geq 25~{\rm GeV}, p_T^b \geq 25~{\rm GeV},
   p_T^\ell \geq 25~{\rm GeV}, \slashed{E}_T^{miss} \geq 25~{\rm GeV}, \\\nonumber
&&\rm |\eta^{jet}|<2.5,  |\eta^b|<2.5, |\eta^\ell|<2.5, \\
&&\rm \Delta R (j j)>0.4,  \Delta R (j b)>0.4, \Delta R (\ell j)>0.4 , \Delta R (\ell b)>0.4 ,
\end{eqnarray}
where $\rm \Delta R = \sqrt{\Delta \Phi^2 + \Delta \eta^2}$ is the separation
in the rapidity-azimuth plane. $\rm p_T^{jet,\ell}$ are the transverse momentum
of jets and leptons and $\rm \slashed{E}_T^{miss}$ is the transverse
missing energy of the neutrino. These general cuts are the basic cuts we apply
in our calculation except special cases where addressed.

\begin{figure}[hbtp]
\centering
\includegraphics[scale=0.8]{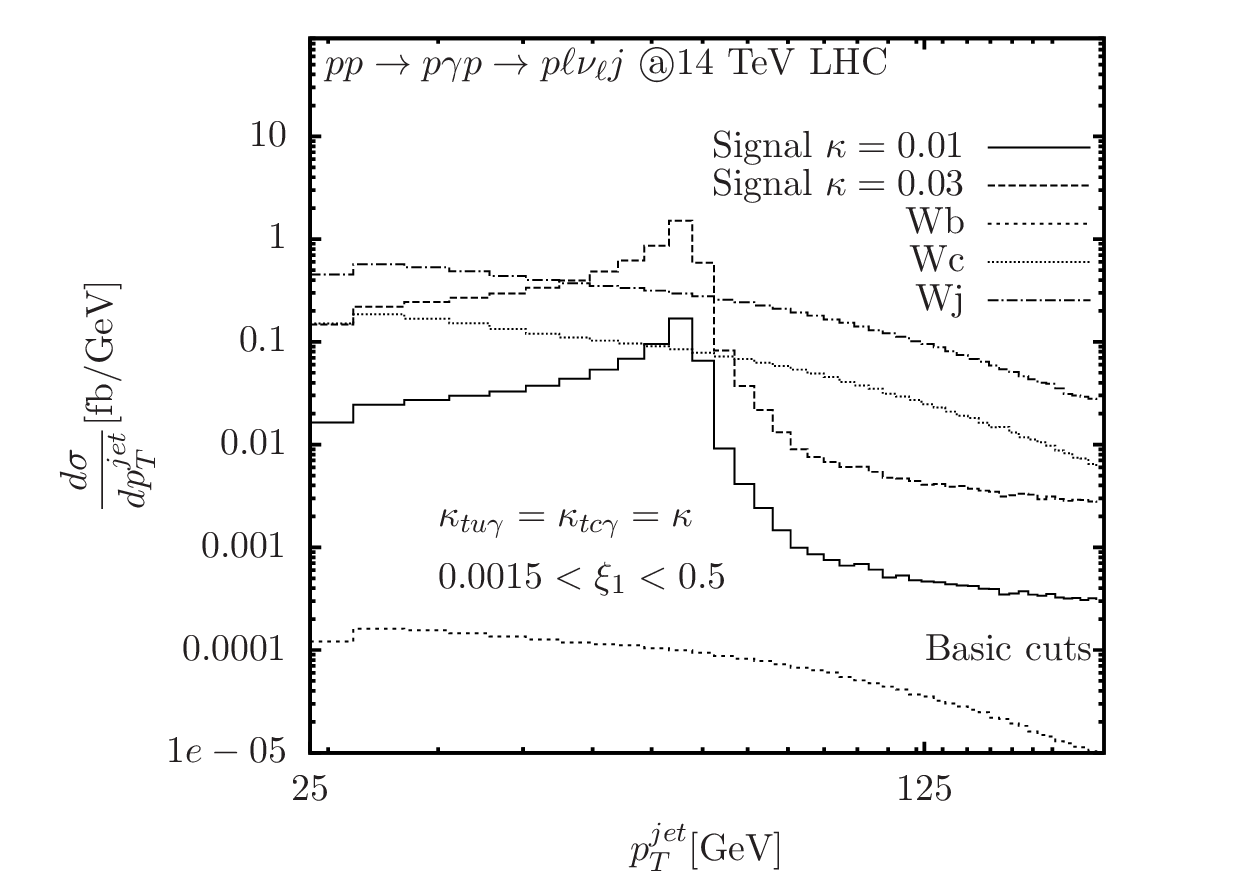}
\caption{\label{fig2}
The transverse momentum distributions for the jet ($\rm p_T^{jet}$) of
$\rm pp\rightarrow p\gamma p\rightarrow pW(\rightarrow\ell\nu_{\ell})j$ ($\rm \ell=e$, $\mu$)
with basic cuts in Eq.(\ref{basiccuts}).
The anomalous coupling is chosen to be $\rm \kappa_{tu\gamma}=\kappa_{tc\gamma}=\kappa=0.01(0.03)$.
The forward detector acceptance is chosen to be $0.0015<\xi_1<0.5$.
The b-tagging efficiency and the rejection factors for the light jets are taken into account.}
\end{figure}

\begin{figure}[hbtp]
\centering
\includegraphics[scale=0.6]{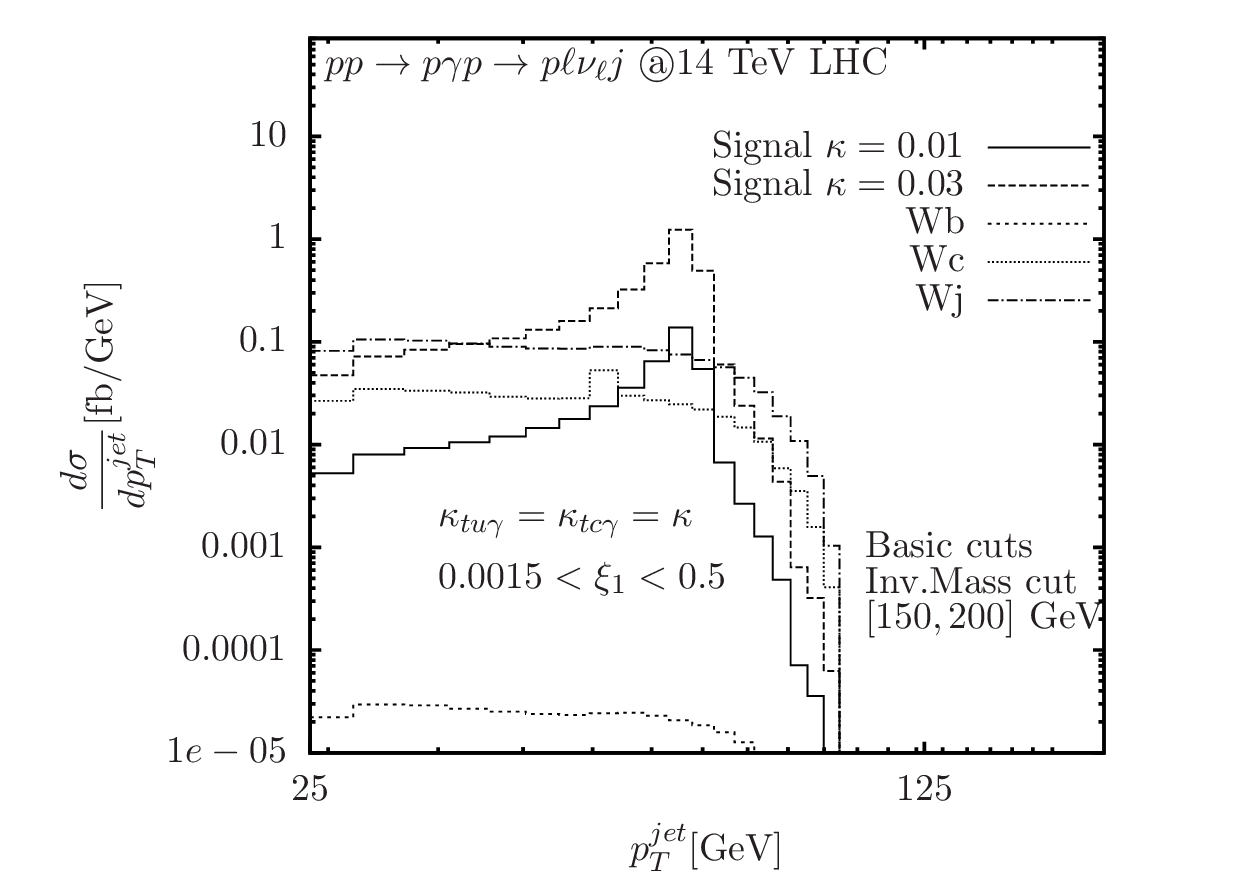}
\includegraphics[scale=0.6]{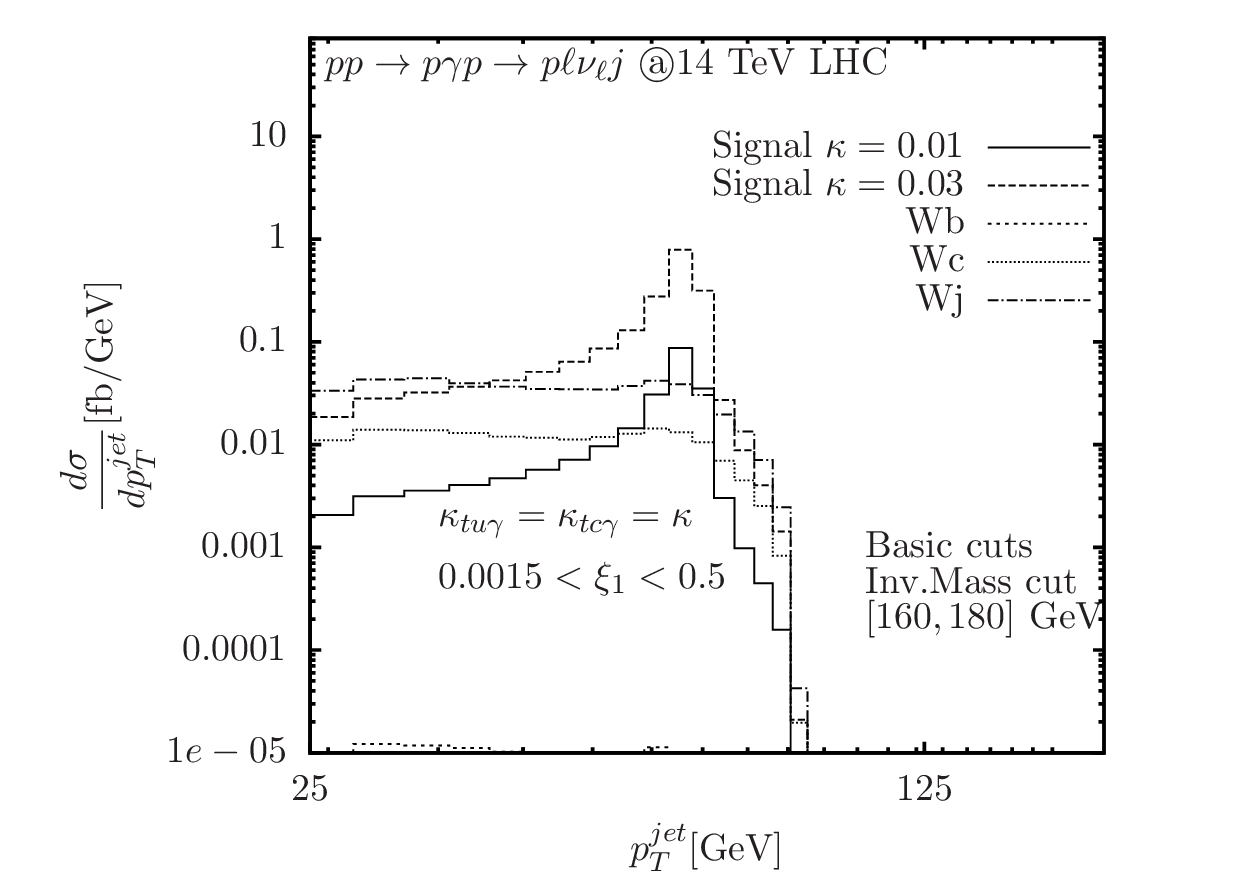}
\caption{\label{fig3}
The transverse momentum distributions for the jet ($\rm p_T^{jet}$) of
$\rm pp\rightarrow p\gamma p\rightarrow pW(\rightarrow\ell\nu_{\ell})j$ ($\rm \ell=e$, $\mu$)
with basic cuts in Eq.(\ref{basiccuts}) plus invariant mass (W+jet)
cut ($\rm 150~{\rm GeV}<M_{Wj}<200~{\rm GeV}$[left panel],
     $\rm 160~{\rm GeV}<M_{Wj}<180~{\rm GeV}$[right panel]).
The anomalous coupling is chosen to be $\rm \kappa_{tu\gamma}=\kappa_{tc\gamma}=\kappa=0.01(0.03)$.
The forward detector acceptance is chosen to be $0.0015<\xi_1<0.5$.
The b-tagging efficiency and the rejection factors for the light jets are taken into account.}
\end{figure}

\begin{figure}[hbtp]
\centering
\includegraphics[scale=0.6]{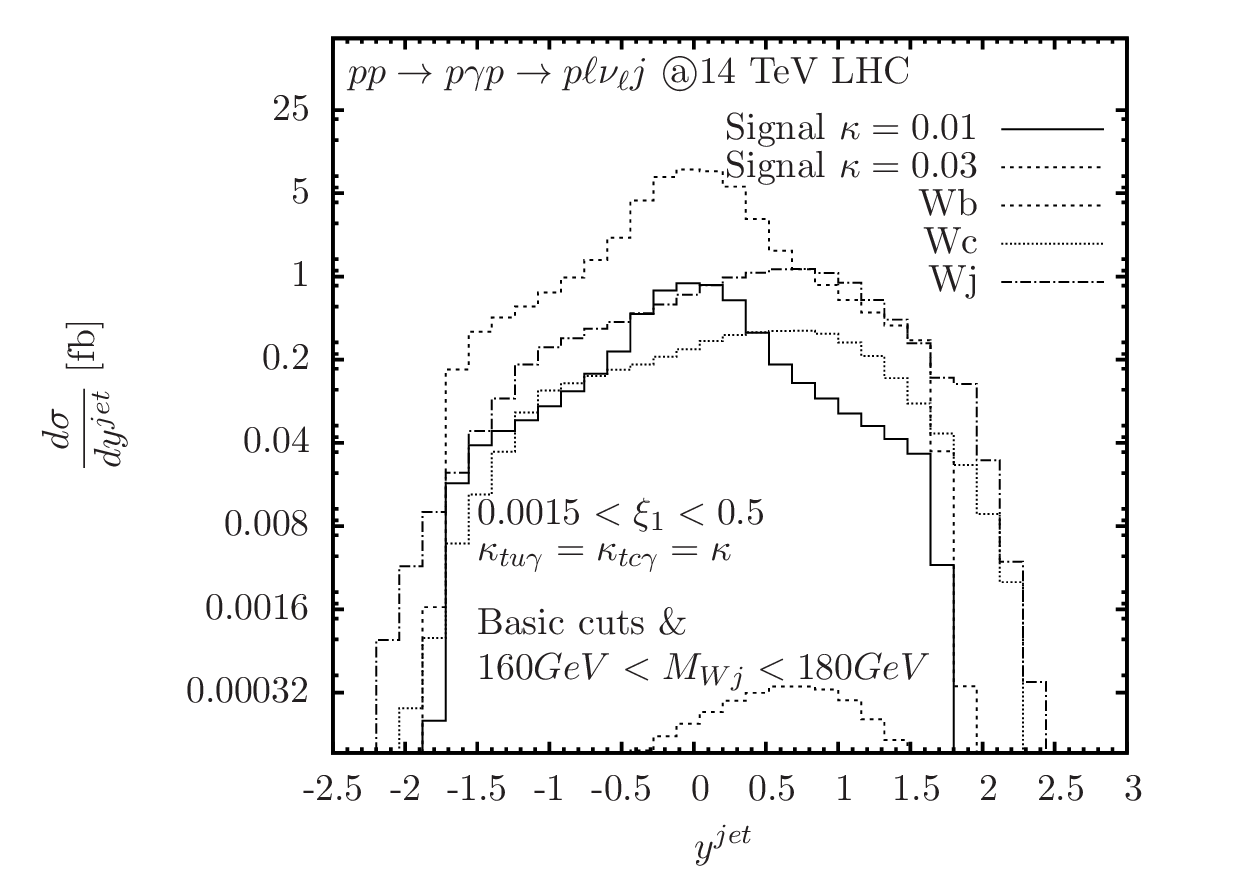}
\includegraphics[scale=0.6]{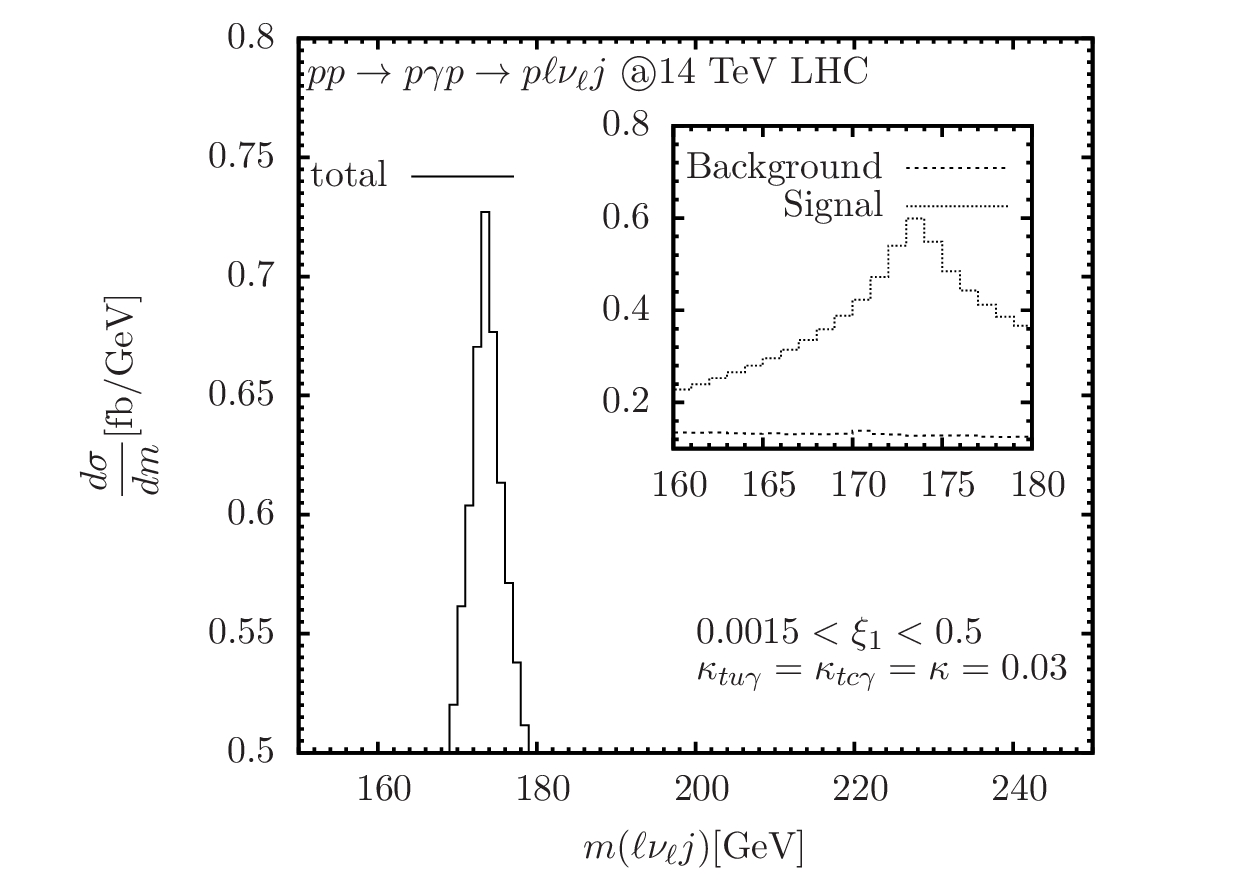}
\caption{\label{fig4}
The pseudorapidity distributions for the jet ($\rm y^{jet}$) of
$\rm pp\rightarrow p\gamma p\rightarrow pW(\rightarrow\ell\nu_{\ell})j$ ($\rm \ell=e$, $\mu$)
with basic cuts in Eq.(\ref{basiccuts}), $\rm p_T^{jet}>35\ GeV$ plus invariant mass (W+jet)
cut ($\rm 160\ GeV<M_{Wj}<180~\ GeV$) [left panel]
and the reconstruction of the top quark mass [right panel].}
\end{figure}

The transverse momentum differential cross sections of the final state
jets ($\rm p_T^{jet}$) are given in Fig.\ref{fig2}. The anomalous coupling
is chosen to be $\rm \kappa_{tu\gamma}=\kappa_{tc\gamma}=\kappa=0.01(0.03)$
and the forward detector acceptance is chosen to be $0.0015<\xi_1<0.5$.
The b-tagging efficiency and the rejection factors for the light jets
and the basic kinematical cuts are taken into account.
In the $\rm p_T$ distribution we can clearly see a resonance in the signal
which correspond to the top quark. In order to improve
the signal to background ratio we can apply an invariant mass cut on the
W-jet system around the top quark mass.
To determine the invariant mass of the W-jet system, we follow Ref.\cite{tqv_ep1,tqv_pp4}
and reconstruct $\rm p_t=p_\ell+p_\nu+p_{b-jet}$.
The transverse momentum of the neutrino can be deduced from the missing transverse momentum.
The longitudinal component of the neutrino momentum is given by
\begin{eqnarray}
\rm  p^\nu_L=\frac{\chi p^\ell_L \pm \sqrt{p^2_\ell(\chi^2-p^2_{T\ell}p^2_{T\nu})}}{p^2_{T\ell}}
\end{eqnarray}
where $\rm \chi=\frac{m^2_W}{2}+p^\ell_T \cdot p^\nu_T$ and $\rm p_L(p_T)$ refer to the longitudinal and
transverse momenta, respectively.
In Fig.\ref{fig3}, the differential
cross sections for signal and background processes are given using invariant mass cut
$\rm 150~{\rm GeV}<M_{Wj}<200~{\rm GeV}$[left panel] and
$\rm 160~{\rm GeV}<M_{Wj}<180~{\rm GeV}$[right panel]
in addition to the basic cuts in Eq.(\ref{basiccuts}).
We see that the invariant mass cut can reduce the W-jet background obviously
while make the signal reduce slightly.
To see how the cross sections for signal and background depend on the final
jet ($\rm p_T^{jet}$) cuts, we present this dependence in Tab.\ref{rqwb_pt5_table}
with the invariant mass cut taken to be $\rm 160~{\rm GeV}<M_{Wj}<180~{\rm GeV}$.
We find that for small value of $\kappa$, for example, $\kappa=0.01$,
larger $\rm p_T^{jet}$ cut can reduce the background cross section essentially
while make the signal reduce slightly. This can be seen directly by comparing
differential cross sections in Fig.\ref{fig2} and Fig.\ref{fig3}.
We also calculate the statistical significance (SS) for the signal and background
on different values of $\rm p_T^{jet}$ cuts in Tab.\ref{rqwb_SS_table}
with the following formula\cite{SSformula}:
\begin{eqnarray}\label{SSformula}
\rm SS=\sqrt{2 [(S+B) log(1+\frac{S}{B}) - S] }
\end{eqnarray}
where S and B are the numbers of signal and background events, respectively.
$\cal L$ presents the luminosity of future 14 TeV LHC.
$\rm (S,B)=\sigma_{(S,B)} \times {\cal L} \times \epsilon$, where $\epsilon$ is the overall
detection efficiency of 0.3 by using this photonproduction channel at the LHC.
We can see for $\kappa=0.01$, statistical significance can be improved
with the $\rm p_T^{jet}$ cuts become larger
while for $\kappa=0.05$, statistical significance is reduced slightly.
In Fig.\ref{fig4}[left panel], we present the pseudorapidity of final state jet.
Parameters and kinematical cuts applied are the same as in the right panel of Fig.\ref{fig3}.
We can find the difference between the signal and the background.
Still, the Wb background is quite small while production of Wc is much larger.
However, both of them are smaller than that of Wj contribution.
We also reconstruct the top quark in Fig.\ref{fig4}[right panel]
with $\rm \kappa_{tu\gamma}=\kappa_{tc\gamma}=\kappa=0.03$.
Dotted, dashed and solid curves present the signal, background and their sum respectively.
Sharp resonance around 173.5 GeV can be reconstructed direct related to the top quark mass.
In the following calculation, we apply $\rm p_T^{jet}>35~{\rm GeV}$
and $\rm 160~{\rm GeV}<M_{Wj}<180~{\rm GeV}$.

\begin{table}
\begin{center}
\begin{tabular}{c c c c c c c c}
\hline\hline
 \multicolumn{8}{c}{ Cross Section dependence on $\rm p_T^{jet}$ Cuts} \\
\multicolumn{1}{c}{$\sigma$(pb)}&& $\rm p_T^{jet}>25\ GeV$ && $\rm p_T^{jet}>35\ GeV$ && $\rm p_T^{jet}>45\ GeV$  \\
\hline
Signal($\kappa=0.01$) &&2.5850$\times 10^{-3}$ &&2.3548$\times 10^{-3}$  &&2.0561$\times 10^{-3}$ &   \\
Signal($\kappa=0.05$) &&0.6461$\times 10^{-1}$ &&0.5888$\times 10^{-1}$  &&0.5141$\times 10^{-1}$&   \\
Wb                    &&0.6678$\times 10^{-6}$ &&0.5323$\times 10^{-6}$  &&0.3788$\times 10^{-6}$ &   \\
Wc                    &&0.7819$\times 10^{-3}$ &&0.6161$\times 10^{-3}$  &&0.4541$\times 10^{-3}$ &   \\
Wj                    &&2.4038$\times 10^{-3}$ &&1.8824$\times 10^{-3}$  &&1.3776$\times 10^{-3}$ &   \\
\hline\hline
\end{tabular}
\end{center}
\caption{\label{rqwb_pt5_table}
Signal and background cross section dependence on $\rm p_T^{jet}$ cuts.
Forward detector acceptance is chosen to be $0.0015<\xi_1<0.5$.
The anomalous couplings $\rm \kappa_{tu\gamma}=\kappa_{tc\gamma}=\kappa=0.01(0.05)$.
Basic cuts, the invariant mass cut $\rm 160~{\rm GeV}<M_{Wj}<180~{\rm GeV}$,
the b-tagging efficiency and the rejection factors for the c, $\rm \bar{c}$ and light jets are taken into account.}
\end{table}

\begin{table}
\begin{center}
\begin{tabular}{c c c c c c c c}
\hline\hline
\multicolumn{8}{c}{ Statistical Significance(SS) dependence on $\rm p_T$ Cuts} \\
\hline
\multicolumn{1}{c}{SS($\kappa=0.01$)}& & $\rm p_T^{jet}>25\ GeV$ && $\rm p_T^{jet}>35\ GeV$ && $\rm p_T^{jet}>45\ GeV$  \\
${\cal L}=2fb^{-1}$ && 1.00608 && 1.02037  && 1.02168  &  \\
${\cal L}=10fb^{-1}$ && 2.24966 && 2.28163  &&  2.28454  &  \\
${\cal L}=200fb^{-1}$ && 10.0608 && 10.2037  && 10.2168  &   \\
\hline
\multicolumn{1}{c}{SS($\kappa=0.05$)}& & $\rm p_T^{jet}>25\ GeV$ && $\rm p_T^{jet}>35\ GeV$ && $\rm p_T^{jet}>45\ GeV$  \\
$\rm {\cal L}=2fb^{-1}$ && 13.0858 && 12.8506 &&  12.3932  & \\
$\rm {\cal L}=10fb^{-1}$ && 29.2606 && 28.7348 &&  27.7121  & \\
\hline\hline
\end{tabular}
\end{center}
\caption{\label{rqwb_SS_table}
Statistical significance(SS) dependence on $\rm p_T^{jet}$ cuts.
Forward detector acceptance is chosen to be $0.0015<\xi_1<0.5$.
The anomalous couplings $\rm \kappa_{tu\gamma}=\kappa_{tc\gamma}=\kappa=0.01(0.05)$.
Basic cuts, the invariant mass cut ($\rm 160~{\rm GeV}<M_{Wj}<180~{\rm GeV}$),
the b-tagging efficiency and the rejection factors for the c, $\rm \bar{c}$ and light jets
as well as the detector simulation effects are taken into account.}
\end{table}

\begin{figure}[hbtp]
\centering
\includegraphics[scale=0.8]{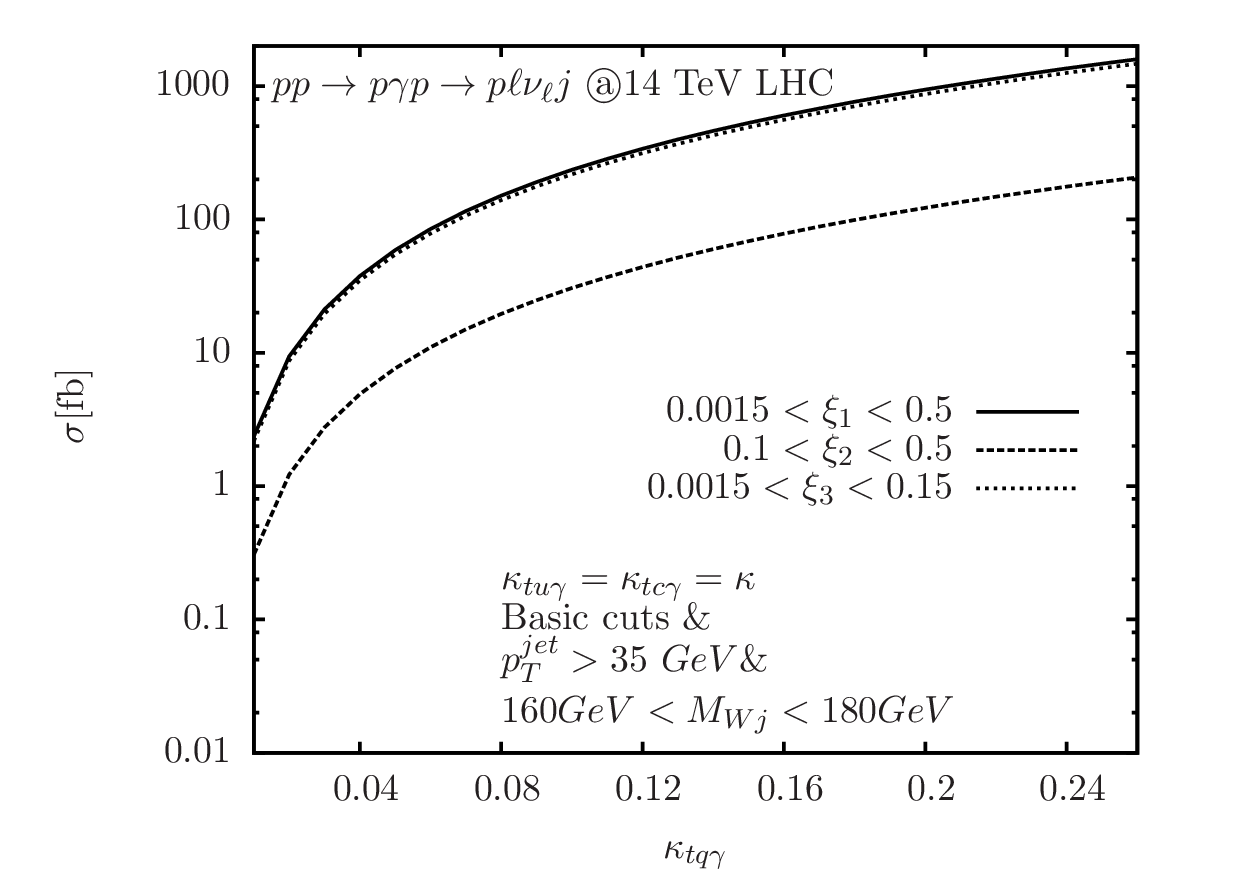}
\caption{\label{rqwb_cross} The total cross sections of signal processes
$\rm pp\rightarrow p\gamma p\rightarrow pW(\rightarrow\ell \nu_{\ell}) b$
as functions of the anomalous $\rm \kappa_{tq\gamma}$ coupling and three forward detector
acceptance regions: $0.0015<\xi_1<0.5$, $0.1<\xi_2<0.5$ and $0.0015<\xi_3<0.15$.}
\end{figure}

In Fig.\ref{rqwb_cross}, we present the signal cross sections of
$\rm pp\rightarrow p\gamma p\rightarrow pW(\rightarrow\ell \nu_{\ell}) b$ ($\rm \ell=e$, $\mu$)
as functions of the anomalous $\rm \kappa_{tq\gamma}$ couplings
and three forward detector acceptance: $0.0015<\xi_1<0.5$, $0.1<\xi_2<0.5$ and $0.0015<\xi_3<0.15$.
Compare different acceptance regions we see that although lines correspond to $\xi_1$, $\xi_2$ and $\xi_3$
have almost the same features, $\xi_1$ and $\xi_3$ do not differ much from each other while both
of them are much larger than cross section of $\xi_2$. We observe from these figures
that cross sections are large for high values of $\rm \kappa_{tq\gamma}$ and are sensitive to
the anomalous couplings as expected. The SM backgrounds for the main reactions are
$\rm \sigma_{B}=2.4985\ fb$ for $0.0015<\xi_1<0.5$,
$\rm \sigma_{B}=0.3311\ fb$ for $0.1<\xi_2<0.5$ and
$\rm \sigma_{B}=2.3117\ fb$ for $0.0015<\xi_3<0.15$.
From this point, we see deviation of the anomalous cross section
from the SM backgrounds are obvious which might detectable from future experiments.

\subsection{Single Top Jet Associated Photoproduction}

\begin{figure}[hbtp]
\vspace{-5cm}
\hspace*{-3cm}
\includegraphics[scale=0.9]{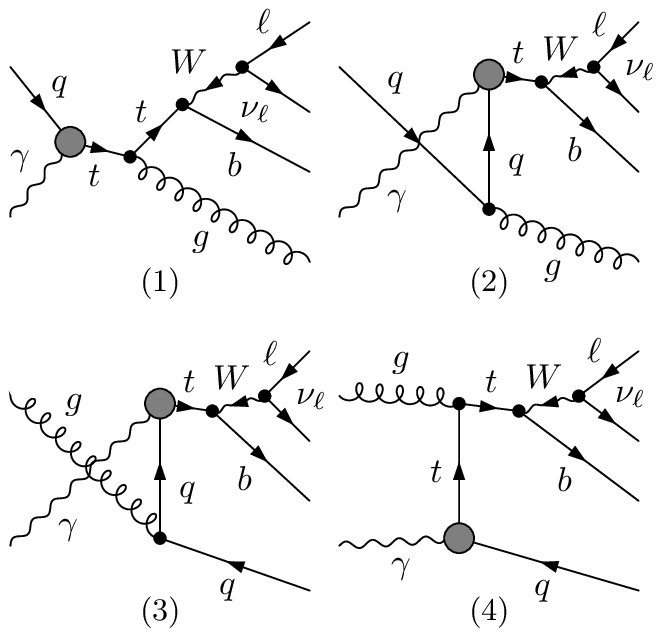}
\vspace{-17cm}
\caption{\label{rqtj} Partonic Feynman diagrams for
$\rm \gamma q/g \rightarrow W(\rightarrow \ell\nu_\ell) b j$ with q=u ,c, j=q, $\rm \bar{q}$, g.
Black blobs represent the anomalous $\rm tq\gamma$ couplings parameterized by Eq.(\ref{lagrangian}).}
\end{figure}

The second single top photoproduction with the anomalous $\rm tq\gamma$ interactions
we examined via the main processes
\begin{eqnarray}\label{processes} \nonumber
&&\rm pp\rightarrow\gamma q      \rightarrow t g    \rightarrow W^+ b g \rightarrow \ell^+ \bar\nu_{\ell} b g  \\\nonumber
&&\rm pp\rightarrow\gamma \bar{q}\rightarrow\bar{t}g\rightarrow W^-\bar{b} g\rightarrow\ell^-\nu_{\ell}\bar{b}g \\\nonumber
&&\rm pp\rightarrow\gamma g\rightarrow t\bar{q}\rightarrow W^+ b\bar{q}\rightarrow\ell^+\bar\nu_{\ell} b\bar{q}\\
&&\rm pp\rightarrow\gamma g\rightarrow\bar{t}q\rightarrow W^-\bar{b}q\rightarrow \ell^- \nu_{\ell} \bar{b} q
\end{eqnarray}
with q=u, c, where we simply refer to these processes as tj productions.
The main reactions include parton level photon-quark collision $\rm \gamma q\rightarrow t g$
and photon-gluon collision $\rm \gamma g\rightarrow t q$.
The motivation for the study of tj process is that:
first, tj associated production is another interesting single top photoproduction
through $\rm \gamma p$ collision at the LHC in addition to the direct single top photoproduction,
both study on them would provide complementary information from one to the other;
second, although an additional particle appear in the final state,
another $\rm \gamma g$ collision mode may also appear.
Since the larger value of gluon parton distribution function, it will be interesting to find out
how this tj channel works to detect the anomalous $\rm tq\gamma$ couplings.
Some Feynman diagrams are shown in Fig.\ref{rqtj}(1-4).
Same as before that the black blobs in these figures
represent the anomalous $\rm tq\gamma$ couplings.
Still we concentrate on the semi-leptonic decay of the single top quark, taking $\rm \ell=e$, $\mu$.
Both the process and its charge-conjugate state are implied.
As can be seen, the studied topology of our signal in this case
therefore give rise to the $\rm \ell j j \slashed E_T$
signature characterized by two jets, one of them tagged as a b-jet, one isolated, either positive or negative,
lepton $\ell^\pm$, and a missing transverse momentum ($\rm \slashed E_T$) from the undetected neutrino.
From this point we can see that the main SM background processes come from mainly two types:
the irreducible background and the reducible ones.

The irreducible background comes from the SM process
$\rm pp\rightarrow p\gamma p\rightarrow W(\rightarrow \ell \nu_\ell)bj$, which yields the identical final state.
In order to get the anomalous $\rm tq\gamma$ coupling effects,
we need to simulate all the signal contributions listed in Eq.(\ref{processes}) precisely
as well as these irreducible backgrounds and their interference.
The total cross section for these reactions thus can be split into three contributions
\begin{eqnarray}
\rm  \sigma=a_0+ a_1 \kappa_{tq\gamma}  + a_2 \kappa_{tq\gamma}^2
\end{eqnarray}
where $\rm a_0$ is the SM prediction, the term $\rm a_1$ linear in $\rm \kappa_{tq\gamma}$
arises from the interference between SM and the anomalous amplitudes, whereas
the quadratic term $\rm a_2$ is the self-interference of the anomalous amplitudes.
Here we still assume $\rm \kappa_{tu\gamma}=\kappa_{tc\gamma}=\kappa$. Indeed, our
results show that the irreducible background without anomalous couplings
(refer to Wbj productions) are small. One reason is the kinematical cuts we
have applied (see bellow) and the other thanks to the small CKM matrix $\rm V_{qq'}$
where q and $\rm q'$ are not the same generation. Here we consider all the possibilities
including the mix generation cases.

Potentially reducible backgrounds come from various other SM processes
that yield different final states which are attributed to the tj signature
due to a misidentification of one or more of the final state objects.
The most important reducible background processes come from light jet faking a b-jet.
Here we include all the Wcj, Wjj productions. The second kind of backgrounds
result from Z-bosons decays to leptons, where one lepton is outside the detector
coverage ($|\eta^\ell| > 2.5$) and fakes missing energy. In this case we consider
all the Zbj, Zcj and Zjj productions. The third kind of possibility comes
from a $\rm W^+W^-j$ production with one W boson decay leptonic and the other W boson decay
hadronic and a jet falls outside of detection. Other kinds of backgrounds result from
$\rm \gamma q/g\rightarrow WZjj'$ where Z boson decay to neutrinos detected as missing energy.
Or single top production like $\rm rq\rightarrow tbq'$, with q, $\rm q'$ present light quarks.
Just like a Wj production with W couples to a top and b quark. And finally decay
to $\rm \ell^\pm\nu_{\ell} j j j'$ with a jet falls outside of detection. These contributions
are quite small and can be safely neglected thus not considered in our calculation.

\begin{table}
\begin{center}
\begin{tabular}{c c}
\hline\hline
Process & Measurable Cross Section[{\rm fb}] \\
\hline
$\rm tj(\kappa=0.01)$&1.1563\\
\hline
$Wbj$& $\sim$ 0.04\\
$Wcj$& $\sim$ 1.43 \\
$Wjj$& $\sim$ 0.89\\
$Zjj$& ${\cal O}(10^{-4})$\\
$WWj$& ${\cal O}(10^{-4})$\\
$WZjj$& $<10^{-5}$\\
\hline\hline
\end{tabular}
\end{center}
\caption{\label{SBcrosssection}
Signal and Background cross sections after the application of the cuts in Eqs.(\ref{basiccuts}) and
$\rm 150~{\rm GeV}< m(\ell \nu_{\ell} j) < 200~{\rm GeV}$.
The b-tagging efficiency and the rejection factors for the c, $\rm \bar{c}$ and light jets are taken into account.}
\end{table}

Tab.\ref{SBcrosssection} summarises the signal and background cross sections after
the application of the basic cuts in Eq.(\ref{basiccuts}).
However, for the tj production, since the extra jet in this case will be in
forward region already, we do not impose the $|\eta|<2.5$ in this calculation.
In addition to the invariant mass cut of the b-jet, the charged lepton and the neutrino system
($\rm m(\ell \nu_{\ell} j)$) close to the top mass has also been included.
So that we can require the final state to be consistent with the $\rm tj(\bar{t}j)$ production.
Since we have two jets in the final state, we require a random one satisfy
$\rm 150~{\rm GeV}< m(\ell \nu_{\ell} j) < 200~{\rm GeV}$ will pass into our selections.
During calculation, we consider all the backgrounds listed
in Tab.\ref{SBcrosssection} except the ones that can be safely omitted.
We can notice in Tab.\ref{rqwb_pt5_table} that the ratio of the dominant Wj
background and the Wc background is about a factor three.
One would expect a similar ratio for the Wjj and Wcj backgrounds in Tab.\ref{SBcrosssection}
or at least the Wjj background should dominant over the Wcj one.
However, it is not the case in Tab.\ref{SBcrosssection}.
Same as the signal, the main reactions for the Wjj and Wcj backgrounds
also include parton level photon-quark collision $\rm \gamma q\rightarrow Wjj(Wcj)$ and
photon-gluon collision $\rm \gamma g\rightarrow Wjj(Wcj)$.
For the contribution from photon-quark collision, before considering the rejection factors
for the c, $\rm \bar{c}$ and light jets, Wjj cross sections are order of 30 times larger than Wcj as expected.
After considering the rejection factors the ratio of these photon-quark collision contribution
for Wjj and Wcj is about a factor three. This is the same as for the ratio of the Wj and Wc backgrounds
as shown in Tab.\ref{rqwb_pt5_table}.
However, dominant contribution for Wjj and Wcj production indeed come from photon-gluon collision.
Though contribution for $\rm \gamma g\rightarrow Wcj$ and $\rm \gamma g\rightarrow Wjj$ are of the same order,
they are much larger than contribution from $\rm \gamma q$ collisions. In this case,
total contribution for Wjj is only order of six times over Wcj. After consider the
rejection factors, we thus get final order of results as shown in Tab.\ref{SBcrosssection}.

To see how the cross sections and statistical significance depend on the $\rm m(\ell \nu_{\ell} j)=M_{Wj}$ cut,
we also require $\rm 160~{\rm GeV}< m(\ell \nu_{\ell} j) < 180~{\rm GeV}$ and
compare it with the former case ($\rm 150~{\rm GeV}< m(\ell \nu_{\ell} j) < 200~{\rm GeV}$)
in Tab.\ref{rqtj_SS_table}.
We see by applying the invariant mass of $\rm 160~{\rm GeV}< m(\ell \nu_{\ell} j) < 180~{\rm GeV}$,
the signal is reduced slightly while the backgrounds can be reduced obviously thus
leading to a better signal over background ratio and higher statistical significance.
The statistical significance for different values of $\cal L$ is presented in the Tab.\ref{rqtj_SS_table}.
In the following calculation we apply $\rm 160~{\rm GeV}< m(\ell \nu_{\ell} j) < 180~{\rm GeV}$.

\begin{table}
\begin{center}
\begin{tabular}{c c c c c }
\hline\hline
 \multicolumn{5}{c}{ Cross Section and Statistical Significance dependence on $\rm M_{Wj}$[GeV] Cuts} \\
\hline
 \multicolumn{1}{c}{$\sigma$(fb)}&& $\rm 150<M_{Wj}<200$&& $160<M_{Wj}<180$ \\	
Signal($\kappa=0.01$) && 1.1563  && 1.1067  \\
Background            && 2.3624  && 1.0543  \\
\hline
SS && && \\
\hline
$\rm {\cal L}=2fb^{-1}$ &&0.5428 && 0.73013 \\
$\rm {\cal L}=10fb^{-1}$ && 1.21391 && 1.63261 \\
$\rm {\cal L}=200fb^{-1}$ && 5.42879 && 7.30126 \\
\hline\hline
\end{tabular}
\end{center}
\caption{\label{rqtj_SS_table}
Signal and background cross section dependence on $\rm M_{Wj}=m(\ell \nu_{\ell} j)$ cuts.
Forward detector acceptance is chosen to be $0.0015<\xi_1<0.5$.
The anomalous couplings $\rm \kappa_{tu\gamma}=\kappa_{tc\gamma}=\kappa=0.01$.
Other cuts include basic cuts in Eq.(\ref{basiccuts}) and
the invariant mass cut $\rm 150~{\rm GeV}<M_{Wj}<200~{\rm GeV}$ ($160~{\rm GeV}<M_{Wj}<180~{\rm GeV}$).
The b-tagging efficiency and the rejection factors for the c, $\rm \bar{c}$ and light jets,
as well as the detector simulation effects are taken into account.}
\end{table}

\begin{figure}[hbtp]
\centering
\includegraphics[scale=0.6]{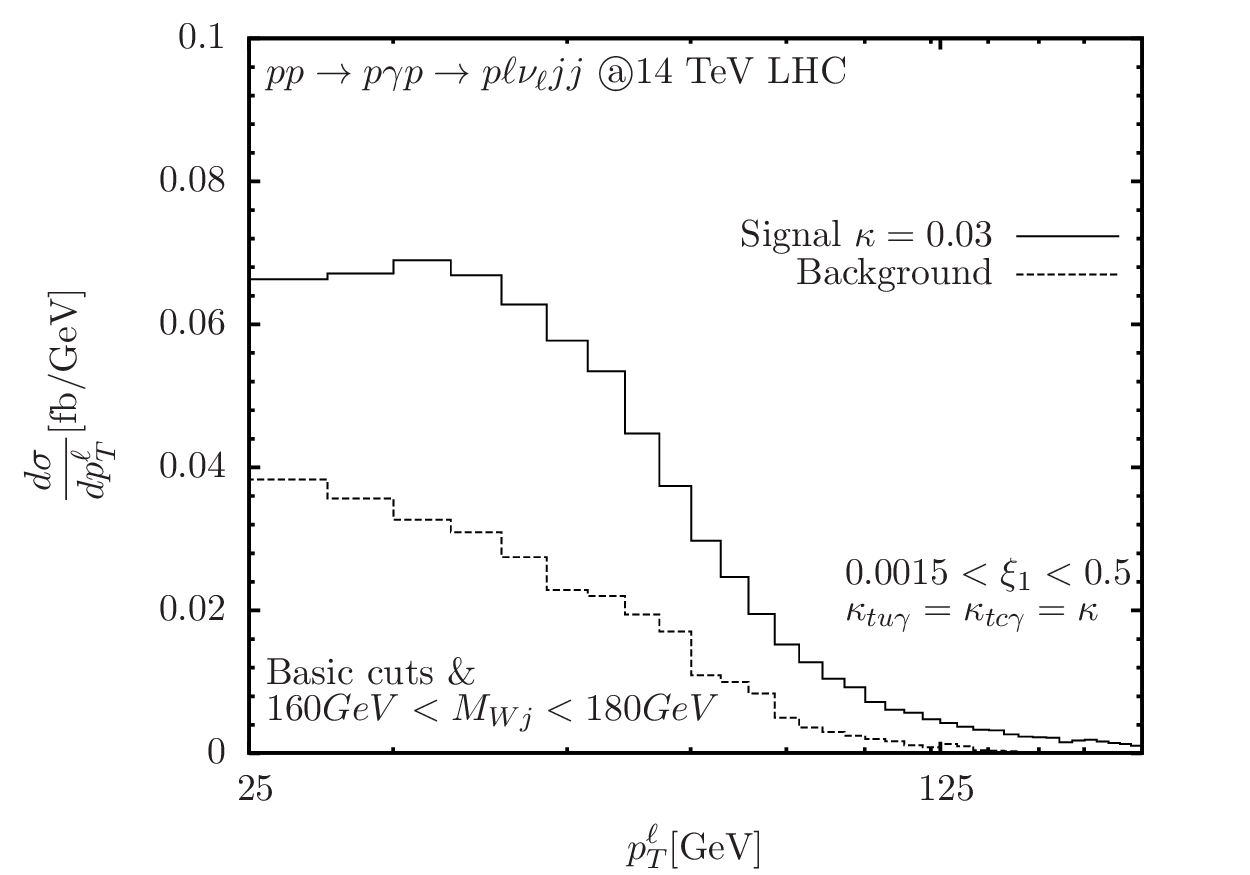}
\includegraphics[scale=0.6]{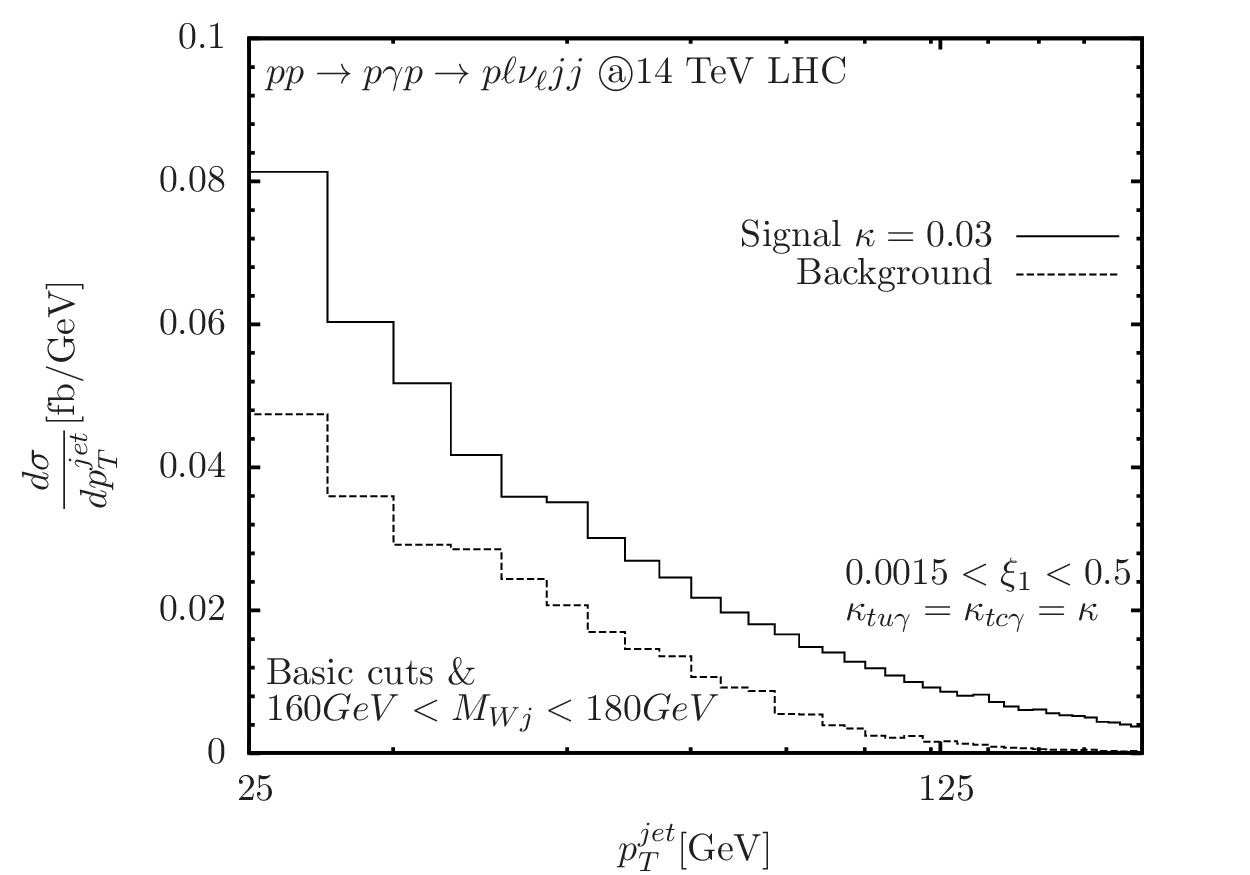}
\includegraphics[scale=0.6]{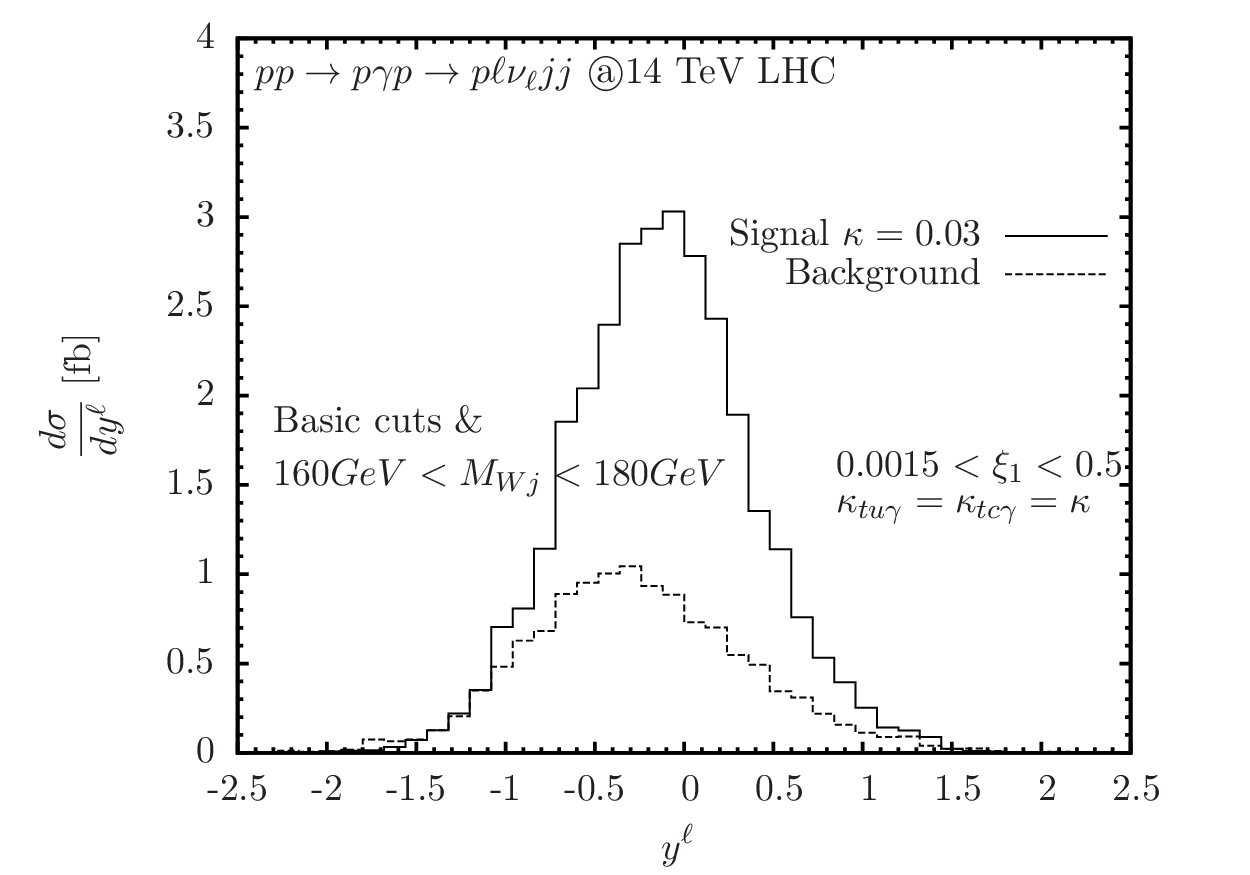}
\includegraphics[scale=0.6]{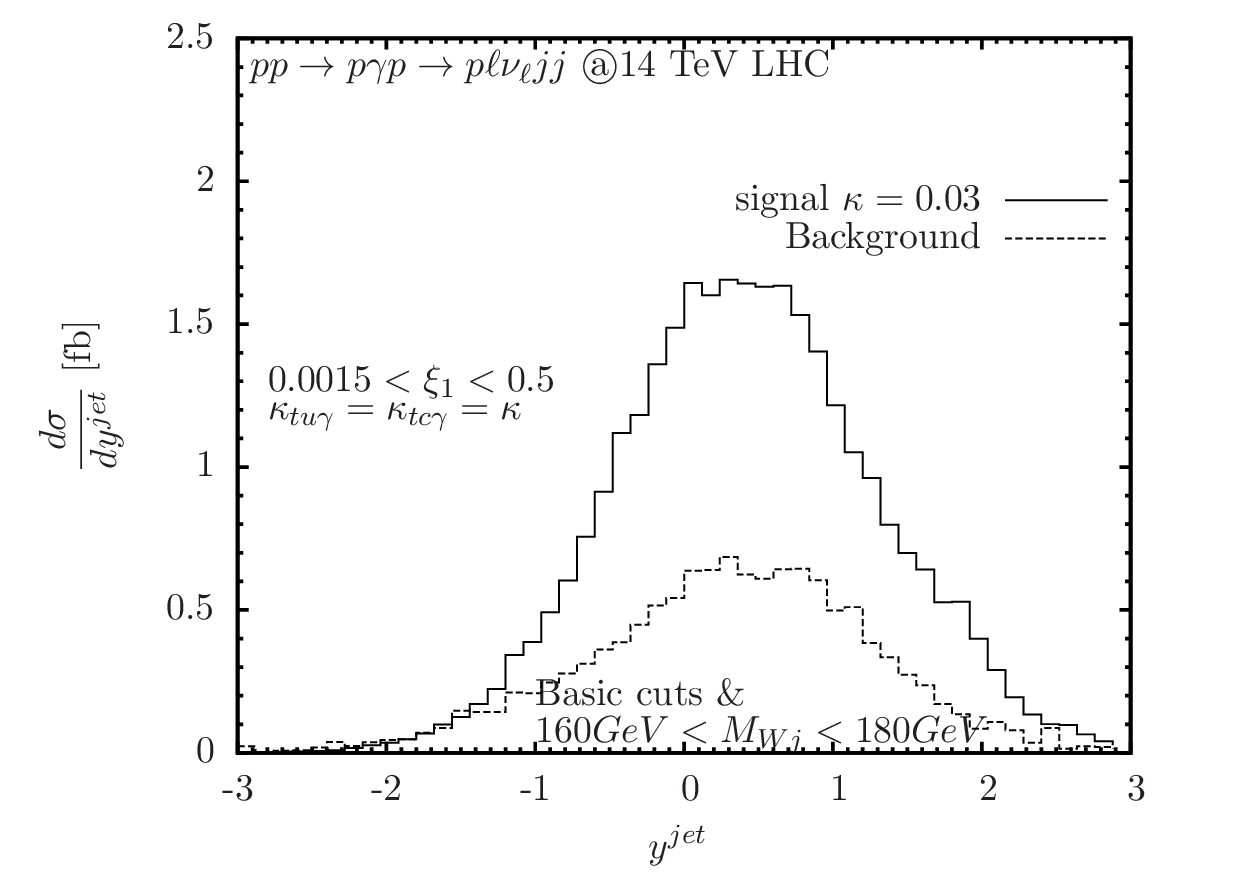}
\includegraphics[scale=0.6]{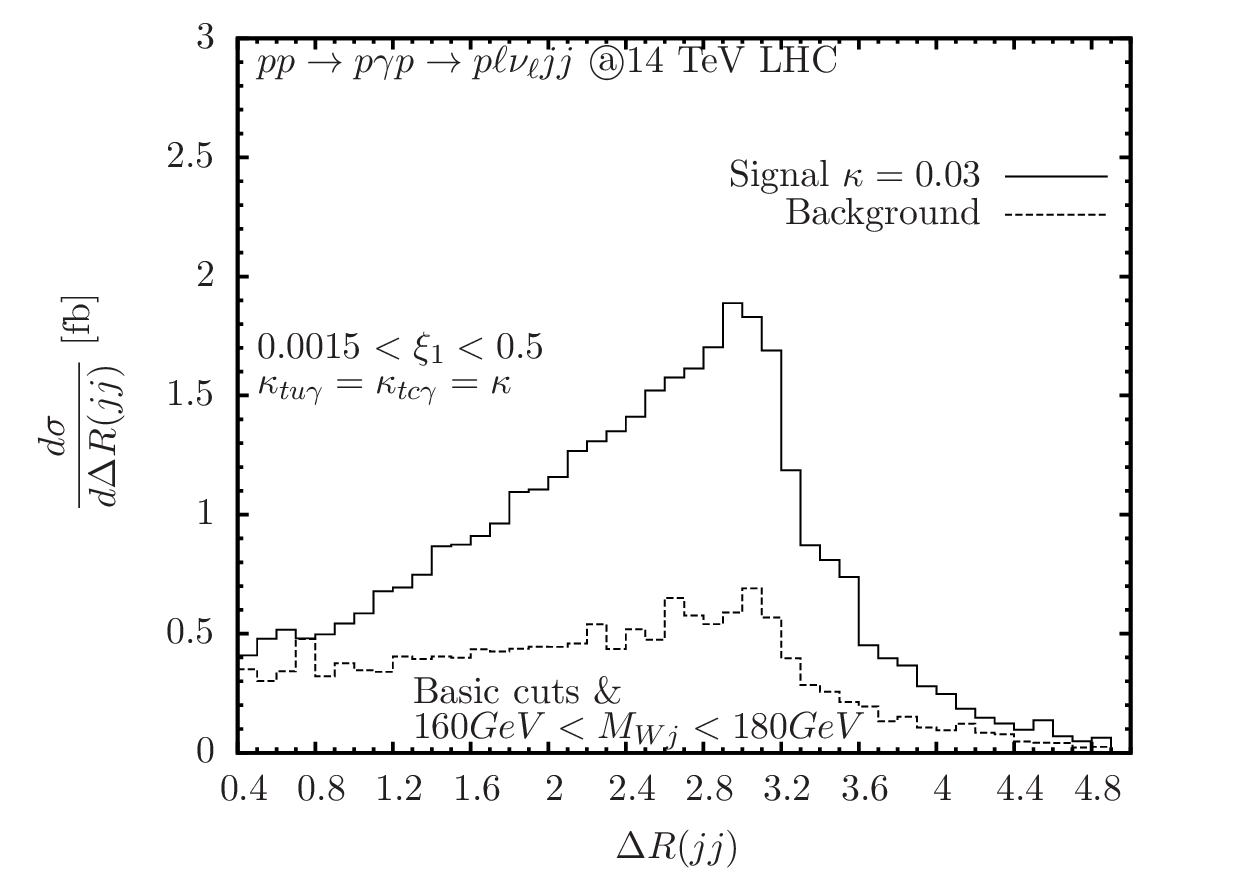}
\includegraphics[scale=0.6]{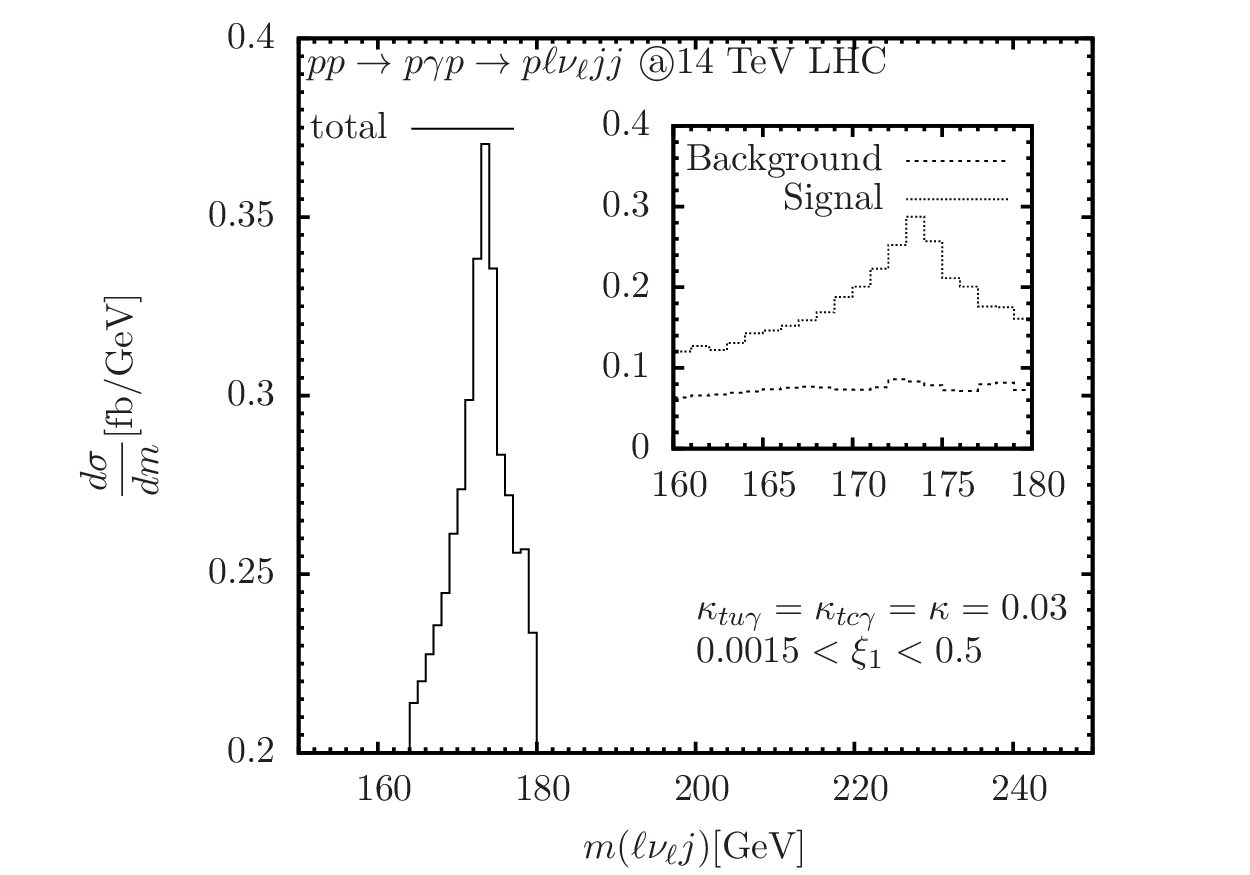}
\caption{\label{rqtj_dis}
The transverse momentum ($\rm p_T^{\ell,jet}$) and rapidity ($\rm y^{\ell,jet}$) distributions for
the charged leptons, $\rm \Delta R(jj)$ distribution of final two jets as well as
the reconstrcution of top mass for
$\rm pp \rightarrow p\gamma p\rightarrow pW(\rightarrow\ell \nu_{\ell}) j j$
($\rm \ell=e$, $\mu$). The anomalous coupling is chosen to be $\rm \kappa_{tq\gamma}=0.03$.
Basic cuts in Eq.(\ref{basiccuts}) and the invariant mass cut $\rm 160~{\rm GeV}<M_{Wj}<180~{\rm GeV}$ are considered.
The b-tagging efficiency and the rejection factors for the c, $\rm \bar{c}$ and light jets are taken into account.}
\end{figure}

In Fig.\ref{rqtj_dis}, we plot the transverse momentum ($\rm p_T^{\ell, jet}$) and rapidity ($\rm y^{\ell,jet}$)
distributions for the charged leptons and the leading jet.
Here two jets are ordered on the basis of their transverse momentum
while the leading one means the one with larger transverse momentum.
$\rm \Delta R(jj)$ distribution of final two jets
as well as the reconstruction of top quark mass are also presented.
The anomalous coupling is chosen to be $\rm \kappa_{tq\gamma}=0.03$.
We see that the background and signal contributions can be well separated.
The $\rm p_T^{\ell,jet}$ distribution can be enhanced at very low $\rm p_T$ regions while reduced at high regions.
$\rm y^{\ell,jet}$ peaks not far from $\rm y=0$ and enhanced obviously in this region.
From $\rm \Delta R$ distribution of final two jets we see the background production
peaks slightly at $\rm \Delta R(jj)=3$ while keep almost flat in the front and middle
while the signal peaked obviously around $\rm \Delta R(jj)=3$. 
Finally, the reconstruction of top quark mass is presented in the last picture of Fig.\ref{rqtj_dis}.
We can clearly see a resonance which corresponds to the top quark with mass of about 173.5 GeV.
Dotted and dashed line present the signal and background,
and are shown in a "zoomed in" range of $\rm 160 \sim 180\ GeV$ in the little figure.
No matter from which one,
signal and background distributions show different features that can be used to separate them.

\begin{figure}[hbtp]
\centering
\includegraphics[scale=0.8]{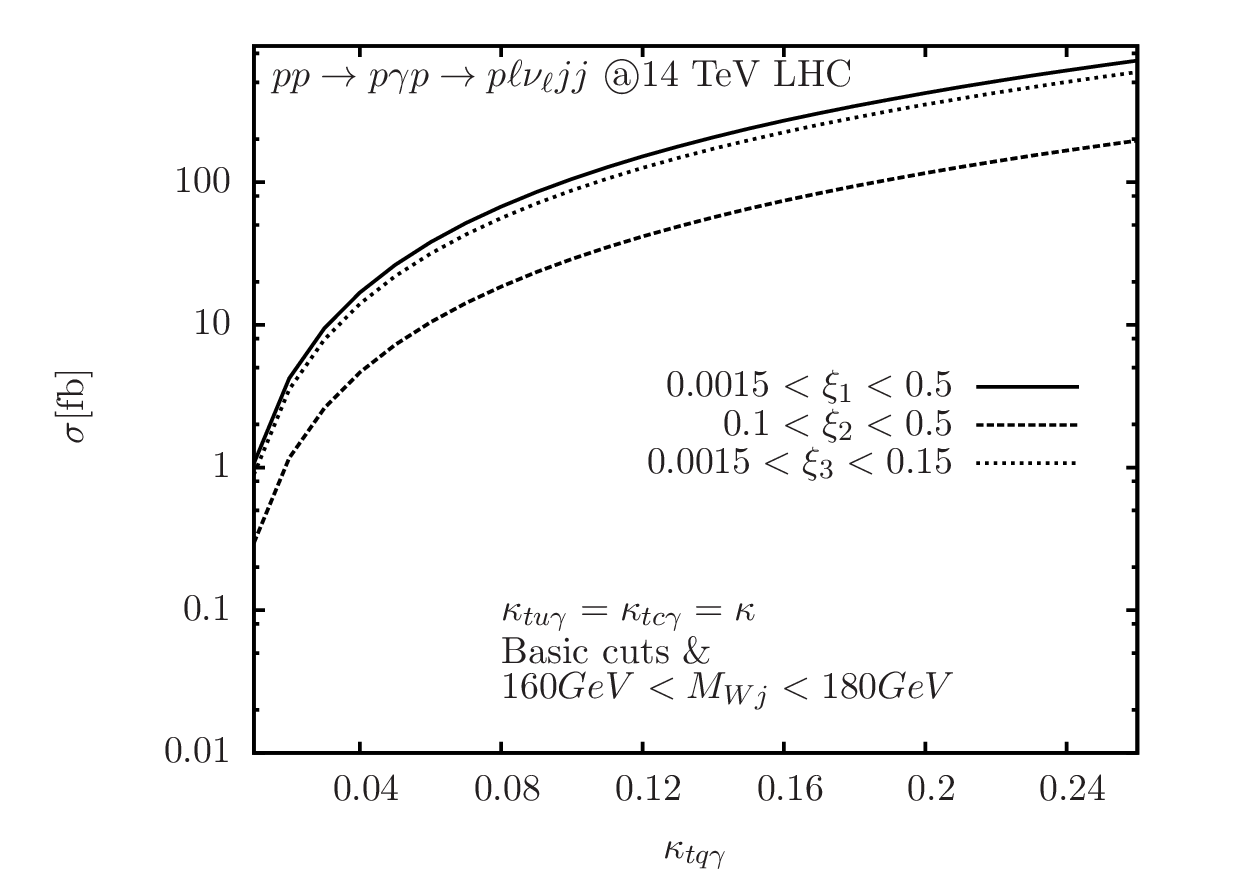}
\caption{\label{rqtj_cross}
The total signal cross sections of $\rm pp \rightarrow p\gamma p\rightarrow pW(\rightarrow \ell \nu_\ell) b j$
($\rm \ell=e$, $\mu$) as functions of the anomalous $\rm \kappa_{tq\gamma}$ coupling and three
forward detector acceptance regions: $0.0015<\xi_1<0.5$, $0.1<\xi_2<0.5$ and $0.0015<\xi_3<0.15$.}
\end{figure}

In Fig.\ref{rqtj_cross}, we present the total signal cross sections of
$\rm pp\rightarrow p\gamma p\rightarrow pW(\rightarrow \ell \nu_\ell) b j$
($\rm \ell=e$, $\mu$) as functions of
the anomalous $\rm \kappa_{tq\gamma}$ coupling and three forward detector
acceptance regions: $0.0015<\xi_1<0.5$, $0.1<\xi_2<0.5$ and $0.0015<\xi_3<0.15$.
The behavior of their productions and their dependence on detector acceptance
are the same as that of $\rm pp\rightarrow p\gamma p\rightarrow pW(\rightarrow\ell \nu_{\ell}) b$:
$\xi_1$ and $\xi_3$ do not differ much from each other while both of them are
much larger than cross section of $\xi_2$.
Here we present the total background cross sections
in Tab.\ref{Bcrosssection} for both
$\rm pp\rightarrow p\gamma p\rightarrow pW(\rightarrow\ell \nu_{\ell}) b$ and
$\rm pp\rightarrow p\gamma p\rightarrow pW(\rightarrow\ell \nu_{\ell}) bj$
which are needed later in the following data analysis.

\begin{table}
\begin{center}
\begin{tabular}{c c c c c c c}
\hline\hline
  \multicolumn{7}{c}{Background Cross Section $\rm \sigma_{B}$[fb]} \\
 \multicolumn{1}{c}{Photoproduction} && $\xi_1$ && $\xi_2$ && $\xi_3$\\	
\hline
$\rm pp\rightarrow p\gamma p\rightarrow pW(\rightarrow\ell \nu_{\ell}) b$   && 2.4985 && 0.3311 && 2.3117\\
$\rm pp\rightarrow p\gamma p\rightarrow pW(\rightarrow\ell \nu_{\ell}) bj$ && 1.0543 && 0.2624 && 0.9311\\
\hline\hline
\end{tabular}
\end{center}
\caption{\label{Bcrosssection}
The total background cross sections
for both $\rm pp\rightarrow p\gamma p\rightarrow pW(\rightarrow\ell \nu_{\ell}) b$ and
$\rm pp\rightarrow p\gamma p\rightarrow pW(\rightarrow\ell \nu_{\ell}) bj$.
Basic cuts, invariant mass cuts, the b-tagging efficiency
and the rejection factors for the c, $\rm \bar{c}$ and light jets are taken into account.}
\end{table}

\section{Bounds for future LHC and the Conclusion}

We follow Ref.\cite{Anomaloustqr} exactly to obtain the sensitivity limits.
Typically, the limits are achieved by assuming the number of
observed events equal to the SM background prediction, $\rm N_{obs}=\sigma_{B} \times {\cal L} \times \epsilon$,
with $\cal L$ for a given integrated luminosity and $\epsilon$ the detection efficiency.
$\rm \sigma_{B}$ is the cross section of SM background prediction.
As can be seen, the SM background events can be less
or larger than 10 for different values of the luminosity
and different types of the detector acceptances.
We thus estimate the sensitivity limits on the anomalous $\rm tq\gamma$ coupling
through these two single top photoproduction channels
by using two different statistical analysis methods
depending on the number of observed events $\rm N_{obs}$.
For $\rm N_{obs} \leq 10$, we employ a Poisson distribution method.
In this case, the upper limits of number of events $\rm N_{up}$
at the 95$\%$ C.L. can be calculated from the formula
\begin{eqnarray}
\rm \Sigma^{N_{obs}}_{k=0} P_{Poisson} (N_{up};k)=1-CL .
\end{eqnarray}
Values for limits candidate $\rm N_{up}$ can be found in Ref.\cite{2012PDG}.
The expected $95\%$ C.L. limits on $\rm \kappa_{tq\gamma}$
can then been calculated by the limits of the observed cross section.
The integrated luminosity $\cal L$ will be taken as a running parameter.
For $\rm N_{obs} > 10$, a chi-square ($\chi^2$) analysis
is performed with the definition
\begin{eqnarray}
\rm  \chi^2 = (\frac{\sigma_{tot}-\sigma_{B}}{\sigma_{B}\delta})^2
\end{eqnarray}
where $\rm \sigma_{tot}$ is the cross section containing new physics effects
and $\rm \delta=\frac{1}{\sqrt{N}}$ is the statistical error
with $\rm N=\sigma_{B} \times {\cal L} \times \epsilon$.
The parameter sensitivity limits on anomalous $\rm tq\gamma$ coupling
as a function of the integrated luminosity can then be obtained.

\begin{figure}[hbtp]
\centering
\includegraphics[scale=0.8]{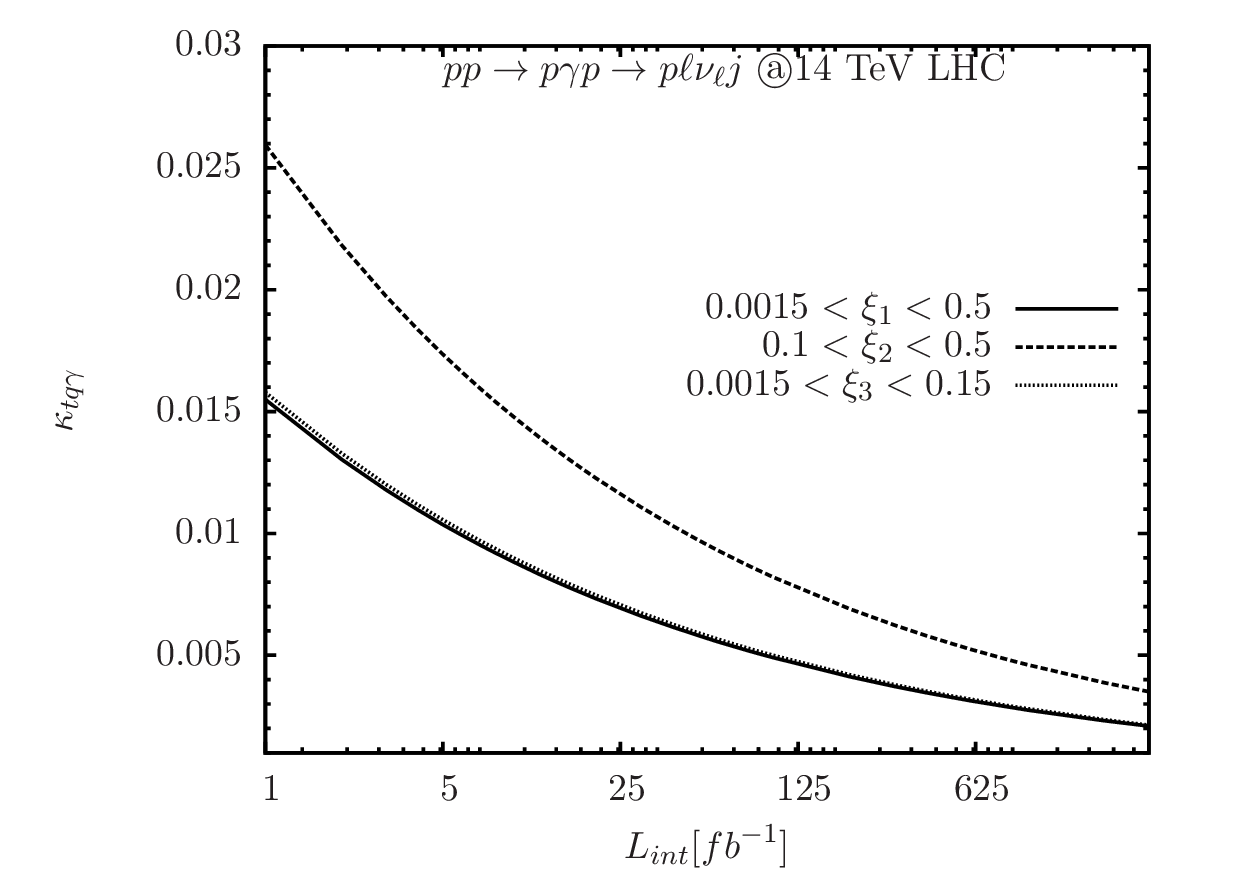}
\caption{\label{rqwb_bound}
$95\%$ C.L. lower bounds for the anomalous $\rm tq\gamma$ couplings
as functions of various integrated luminosity and forward detector acceptances of
$0.0015<\xi_1<0.5$, $0.1<\xi_2<0.5$ and $0.0015<\xi_3<0.15$.
Bounds obtained by using channel $\rm pp\rightarrow p\gamma p\rightarrow pW(\rightarrow\ell \nu_{\ell}) b$.}
\end{figure}

\begin{figure}[hbtp]
\centering
\includegraphics[scale=0.8]{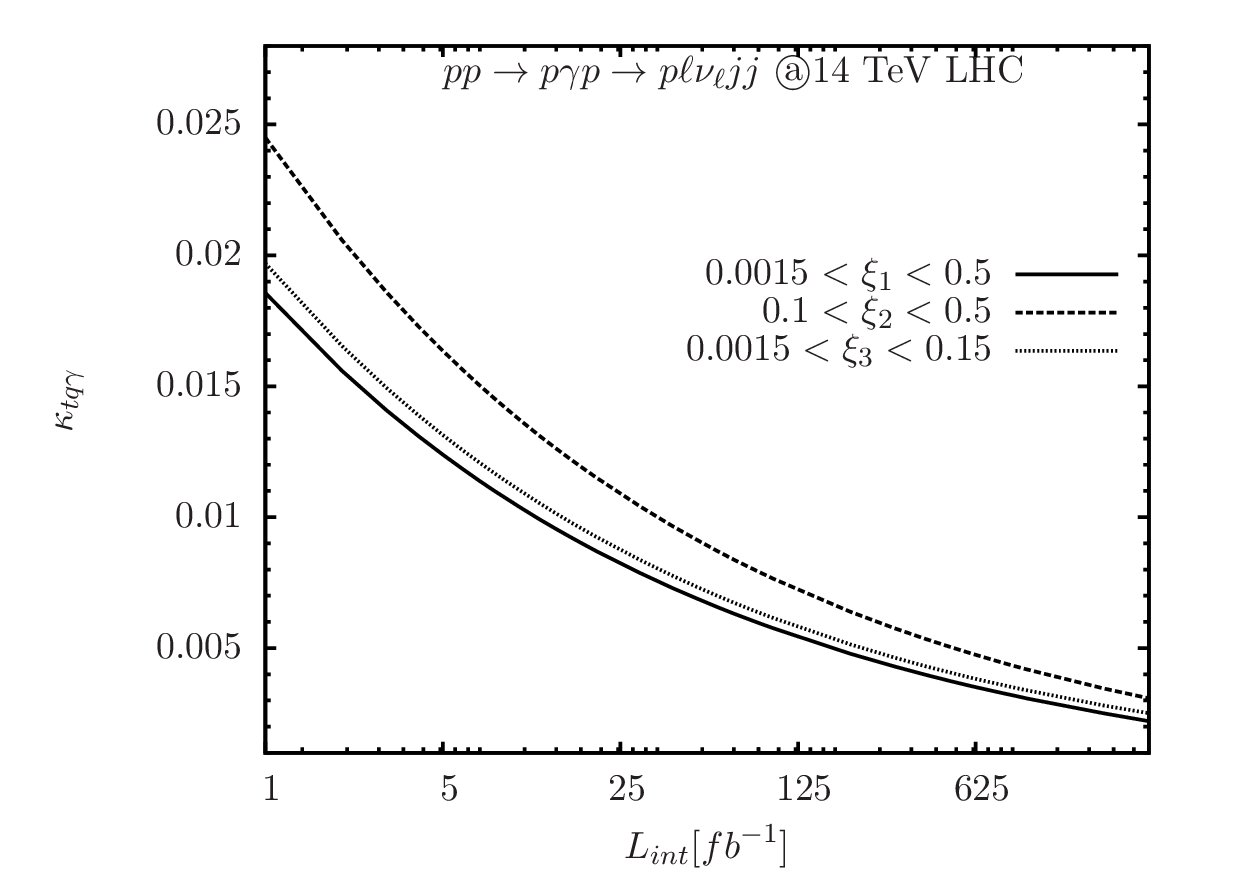}
\caption{\label{rqtj_bound}
$95\%$ C.L. lower bounds for the anomalous $\rm tq\gamma$ couplings
as functions of various integrated luminosity and forward detector acceptances of
$0.0015<\xi_1<0.5$, $0.1<\xi_2<0.5$ and $0.0015<\xi_3<0.15$.
Bounds obtained by using channel $\rm pp\rightarrow p\gamma p\rightarrow pW(\rightarrow\ell \nu_{\ell}) bj$.}
\end{figure}

We present the $95\%$ C.L. sensitivity limits on the anomalous $\rm tq\gamma$ couplings
as functions of various integrated luminosity and forward detector acceptances of
$0.0015<\xi_1<0.5$, $0.1<\xi_2<0.5$ and $0.0015<\xi_3<0.15$
in Fig.\ref{rqwb_bound} by using channel
$\rm pp\rightarrow p\gamma p\rightarrow pW(\rightarrow\ell \nu_{\ell}) b$
and in Fig.\ref{rqtj_bound} by using channel
$\rm pp\rightarrow p\gamma p\rightarrow pW(\rightarrow\ell \nu_{\ell}) bj$.
Choosing which statistical analysis method depends on the number of observed events.
Difference of our final results from Ref.\cite{Anomaloustqr}
mainly due to the different choice of kinematical cuts and the W-jet background simulations.
We recalculate the process of $\rm pp\rightarrow p\gamma p\rightarrow pWb$ in Ref.\cite{Anomaloustqr}
and get the same results following their discussions. This can be a check for both calculations.
By applying the input parameters listed above and kinematical cuts, i.e., $\rm p^{jet}_T>35\ GeV$
for leading single top channel and the invariant mass cut
$\rm 160\ GeV<M_{Wj}<180\ GeV$ for both channels, our results show that:
for the typical detector acceptance $0.0015<\xi_1<0.5$, $0.1<\xi_2<0.5$ and $0.0015<\xi_3<0.15$
with a luminosity of 2 $\rm \rm{fb}^{-1}$  at the future LHC,
the lower bounds of $\rm \kappa_{tq\gamma}$
through leading single top channel (single top jet channel) are
0.0130 (0.0156), 0.0218 (0.0206) and 0.0133 (0.01655), respectively,
correspond to $\rm Br(t\rightarrow q\gamma)\sim 3\times 10^{-5}$.
With a luminosity of 200 $\rm fb^{-1}$, the lower bounds of $\rm \kappa_{tq\gamma}$ are
0.0041 (0.0048), 0.0069 (0.0064) and 0.0042 (0.0051), respectively,
correspond to $\rm Br(t\rightarrow q\gamma)\sim 4\times 10^{-6}$,
see in Tab.\ref{table_bounds} for more details.
We find that for the typical detector acceptance $0.0015<\xi_1<0.5$ and $0.0015<\xi_3<0.15$,
leading single top photoproduction is the better channel to test anomalous $\rm tq\gamma$ couplings
than single top jet channel. While for $0.1<\xi_2<0.5$, single top jet channel becomes better.
Compare these two single top photoproduction processes
$\rm pp\rightarrow p\gamma p\rightarrow pW(\rightarrow\ell \nu_{\ell}) b$
and $\rm pp\rightarrow p\gamma p\rightarrow pW(\rightarrow\ell \nu_{\ell}) bj$,
both channels can be used to test the anomalous $\rm tq\gamma$ couplings.
These parameter limits (bounds) are also comparable with the other
phenomenological studies\cite{tqv_ep2,tqv_ep3,tqv_ep4, Anomaloustqr, tqr_limit1, tqr_limit2}
and much better than
the constraints from experiments\cite{tqr_CDF, tqr_ZEUS, tqr_H1, anomalous_LHC}.
Notice that in Fig.\ref{rqwb_bound} and Fig.\ref{rqtj_bound},
we also present the bounds obtained when the luminosity become
larger than $\rm 200 fb^{-1}$, see, up to $\rm 1000 fb^{-1}$.
However, we should mention here that as the luminosity become larger,
identify the signal under the high pileup running conditions will be challenge:
the hadronic background of multiple pp interactions will be so large that any $\rm \rm \gamma p$
process will be completely swamped. This can be a drawback of $\rm \gamma p$ productions.
In this case a more detailed study on the experimental effects, i.e.,
pileup rejection factors, should also be considered.
These will and might significantly reduce the constraints on the bounds obtained.
However, in our phenomenological study, we keep all the results up to high luminosity
with the discussion been focused only up to $\rm 200 fb^{-1}$.
Full detector simulation is beyond the scope of this analysis.

\begin{table}
\begin{center}
\begin{tabular}{c c c c c c c c}
\hline\hline
 \multicolumn{8}{c}{$95\%$ C.L. lower bounds for the anomalous $\rm tq\gamma$ couplings $\rm \kappa_{tq\gamma}$} \\
 \hline
& \multicolumn{3}{c}{$\rm pp\rightarrow p\gamma p\rightarrow pW(\rightarrow\ell \nu_{\ell}) b$} & &
  \multicolumn{3}{c}{$\rm pp\rightarrow p\gamma p\rightarrow pW(\rightarrow\ell \nu_{\ell}) bj$} \\
${\cal L} [\rm fb^{-1}]$ & $\xi_1$ &$\xi_2$ &$\xi_3$ && $\xi_1$ &$\xi_2$ &$\xi_3$  \\	
\hline
2    &  0.0130  & 0.0218 & 0.0133 && 0.0156  & 0.0206 & 0.01655 \\
200  & 0.0041&	0.0069	& 0.0042 && 0.0048& 0.0064 &	0.0051 \\
\hline\hline
\end{tabular}
\end{center}
\caption{\label{table_bounds}
$95\%$ C.L. lower bounds for the anomalous $\rm tq\gamma$ couplings
as functions of various integrated luminosity and forward detector acceptances of
$0.0015<\xi_1<0.5$, $0.1<\xi_2<0.5$ and $0.0015<\xi_3<0.15$.
Bounds obtained by using channel $\rm pp\rightarrow p\gamma p\rightarrow pW(\rightarrow\ell \nu_{\ell}) b$
and channel $\rm pp\rightarrow p\gamma p\rightarrow pW(\rightarrow\ell \nu_{\ell}) bj$.}
\end{table}

\vskip 5mm
\section{Summary}

In this work, we examine the anomalous $\rm tq\gamma$ (q=u, c) coupling
through photon-produced leading single top production
and single top jet associated production through the main reaction
$\rm pp\rightarrow p\gamma p \rightarrow pt\rightarrow pW(\rightarrow\ell \nu_\ell) b+X$
and $\rm pp\rightarrow p\gamma p\rightarrow ptj\rightarrow pW(\rightarrow\ell \nu_\ell) bj+X$
assuming a typical LHC multipurpose forward detectors in a model independent
effective lagrangian approach. Full effects of the top quark leptonic decay modes
($\rm t\rightarrow Wb \rightarrow \ell \nu_{\ell} b$, with $\rm \ell=e$, $\mu$) are taken into account.
We have employed Equivalent Photon Approximation (EPA) for the incoming photon beams
and performed detailed analysis for various forward detector acceptances ($\xi$).
We analyse their impacts on both the total cross sections and some key distributions.
The full background analysis are considered.
Finally, we present the $95\%$ C.L. sensitivity limits on the anomalous
$\rm tq\gamma$ couplings as functions of different integrated luminosity
and forward detector acceptances through both channels.
With our input parameters and kinematical cuts, results show that:
for the typical detector acceptance $0.0015<\xi_1<0.5$, $0.1<\xi_2<0.5$
and $0.0015<\xi_3<0.15$ with a luminosity of 2 $\rm{fb}^{-1}$,
the lower bounds of $\rm \kappa_{tq\gamma}$
through leading single top channel (single top jet channel) are
0.0130 (0.0156), 0.0218 (0.0206) and 0.0133 (0.01655), respectively,
correspond to $\rm Br(t\rightarrow q\gamma)\sim 3\times 10^{-5}$.
With a luminosity of 200 $\rm fb^{-1}$, the lower bounds of $\rm \kappa_{tq\gamma}$ are
0.0041 (0.0048), 0.0069 (0.0064) and 0.0042 (0.0051), respectively,
correspond to $\rm Br(t\rightarrow q\gamma)\sim 4\times 10^{-6}$.
We see that for the typical detector acceptance $0.0015<\xi_1<0.5$ and $0.0015<\xi_3<0.15$,
leading single top photoproduction is the better channel to test anomalous $\rm tq\gamma$ couplings
than single top jet channel. While for $0.1<\xi_2<0.5$, single top jet channel becomes better.
We conclude that both channels can be used to detect such anomalous $\rm tq\gamma$ couplings
and the detection sensitivity on $\rm \kappa_{tq\gamma}$ is obtained.

\section*{Acknowledgments} \hspace{5mm}
The author thanks Dr. Inanc Sahin and Fawzi Boudjema for their kindness to provide invaluable comments,
thanks Dr. RenYou Zhang, ChongXing Yue, ShouShan Bao, Shuo Yang for useful discussions.
Project supported by the National Natural Science Foundation of China (Grant No. 11205070),
by Shandong Province Natural Science Foundation (Grant No. ZR2012AQ017)
and by the Fundamental Research Funds for the Central Universities (No. DUT13RC(3)30).


\vspace{1.0cm}
\begin{thebibliography}{99}

\bibitem{tqv_SMTHDM1}
B. Grzadkowski, J. F. Gunion, P. Krawczyk,
{\it Neutral current flavor changing decays for the Z boson
and the top quark in two Higgs doublet models},
Phys. Lett. B 268 (1991) 106-111.

\bibitem{tqv_SMTHDM2}
G. Eilam, J. L. Hewett, A. Soni,
{\it Rare decays of the top quark in the standard and two Higgs doublet models},
Phys. Rev. D 44 (1991) 1473-1484, Erratum-ibid. Phys. Rev. D 59 (1999) 039901.

\bibitem{tqv_THDM1}
Santi Bejar, Jaume Guasch, Joan Sola, {\it Loop induced flavor changing neutral
decays of the top quark in a general two Higgs doublet model},
Nucl. Phys. B 600 (2001) 21-38, [arXiv:hep-ph/0011091].

\bibitem{tqv_THDM2}
Itzhak Baum, Gad Eilam, Shaouly Bar-Shalom,
{\it Scalar FCNC and rare top decays in a two Higgs doublet model "for the top"},
Phys. Rev. D 77 (2008) 113008, [arXiv:0802.2622].

\bibitem{tqv_THDM3}
J. L. Diaz-Cruz, M. A. Perez, G. Tavares-Velasco, J . J. Toscano,
{\it Testing flavor-changing neutral currents in the rare decays $t\rightarrow cV_iV_j$}
Phys. Rev. D 60 (1999) 115014, [arXiv:hep-ph/9903299].

\bibitem{tqv_TC}
Xuelei Wang, Gongru Lu, Jinmin Yang, Zhenjun Xiao, Chongxing Yue, Yimin Zhang,
{\it Rare decays of the top quark in the one generation technicolor model},
Phys. Rev. D 50 (1994) 5781-5786;
Gongru Lu, Furong Yin, Xuelei Wang, Lingde Wan,
{\it Rare top quark decays $t\rightarrow cV$ in the top-color-assisted technicolor model},
Phys. Rev. D 68 (2003) 015002, [arXiv:hep-ph/0303122];
Junjie Cao, Guoli Liu, Jin Min Yang, Huanjun Zhang,
{\it Top-quark FCNC Productions at LHC in Topcolor-assisted Technicolor Model},
Phys. Rev. D 76 (2007) 014004, [arXiv:hep-ph/0703308].

\bibitem{tqv_quarksinglet}
F. del Aguilar, J. A. Aguilar-Saavedra, R. Miquel,
{\it Constraints on top couplings in models with exotic quarks},
Phys. Rev. Lett. 82 (1999) 1628-1631, [arXiv:hep-ph/9808400];
J. A. Aguilar-Saavedra, {\it Effects of mixing with quark singlets},
Phys. Rev. D 67 (2003) 035003, Erratum-ibid. D 69 (2004) 099901, [arXiv:hep-ph/0210112];
J. A. Aguilar-Saavedra, B.M. Nobre,
{\it Rare top decays $t\rightarrow c\gamma$, $t\rightarrow cg$ and CKM unitarity},
Phys. Lett. B 553 (2003) 251-260, [arXiv:hep-ph/0210360].

\bibitem{tqv_MSSM1}
Chong Sheng Li, R. J. Oakes, Jin Min Yang,
{\it Rare decays of the top quark in the minimal supersymmetric model},
Phys. Rev. D 49 (1994) 293298, Erratum-ibid. Phys. Rev. D 56 (1997) 3156;
Jian Jun Liu, Chong Sheng Li, Li Lin Yang, Li Gang Jin,
{\it $t \rightarrow cV$ via SUSY FCNC couplings in the unconstrained MSSM},
Phys. Lett. B 599 (2004) 92-101.

\bibitem{tqv_MSSM2}
G. Couture, C. Hamzaoui, H. K$\ddot{o}$nig,
{\it Flavor-changing top quark decay within the minimal supersymmetric standard model},
Phys. Rev. D 52 (1995) 1713-1716;
G. Couture, M. Frank, H. K$\ddot{o}$nig,
{\it Supersymmetric QCD flavor-changing top quark decay}, Phys. Rev. D 56 (1997) 42134218.

\bibitem{tqv_MSSM3}
J.L. Lopez, D.V. Nanopoulos, R. Rangarajan,
{\it New supersymmetric contributions to $t\rightarrow cV$}, Phys. Rev. D 56 (1997) 31003106.

\bibitem{tqv_MSSM4}
G.M. de Divitiis, R. Petronzio, L. Silvestrini,
{\it Flavor changing top decays in supersymmetric extensions of the standard model},
Nucl. Phys. B 504 (1997) 45-60.

\bibitem{tqv_MSSM5}
Jaume Guasch, Joan Sola, {\it FCNC top quark decay in the MSSM:
a door to SUSY physics in high luminosity colliders?},
Nucl. Phys. B 562 (1999) 3-28, [arXiv:hep-ph/9906268];
Santi Bejar, Jaume Guasch, Joan Sola,
{\it FCNC top quark decays beyond the Standard Model}, [arXiv:hep-ph/0101294].

\bibitem{tqv_MSSM6}
Jin Min Yang, Bing-Lin Young, X. Zhang,
{\it Flavor-changing top quark decays in R-parity-violating supersymmetric models},
Phys. Rev. D 58 (1998) 055001.

\bibitem{tqv_MSSM7}
D. Delepine, S. Khalil, {\it Top flavor violating decays in general supersymmetric models},
Phys.Lett. B599 (2004) 62-74.

\bibitem{tqv_MSSM8}
Mariana Frank, Ismail Turan,
{\it Rare decay of the top $t \rightarrow c l^+ l^-$ and single top production at ILC},
Phys.Rev. D74 (2006) 073014.

\bibitem{tqv_warpED}
Kaustubh Agashe, Gilad Perez, Amarjit Soni,
{\it Flavor structure of warped extra dimension models},
Phys.Rev. D71 (2005) 016002, [arXiv:hep-ph/0408134];
{\it Collider Signals of Top Quark Flavor Violation from a Warped Extra Dimension},
Phys.Rev. D75 (2007) 015002, [arXiv:hep-ph/0606293].

\bibitem{tqv_LRSUSY}
Mariana Frank, Ismail Turan,
{\it $t\rightarrow cg, c\gamma, cZ$ in the Left-Right Supersymmetric Model},
Phys. Rev. D 72 (2005) 035008, [arXiv:hep-ph/0506197].

\bibitem{tqv_ILC1}
David Atwood, Laura Reina, Amarjit Soni,
{\it Flavor changing neutral scalar currents at $\mu^+\mu^-$ colliders},
Phys. Rev. Lett. 75 (1995) 3800, [arXiv:hep-ph/9507416];
David Atwood, Laura Reina, Amarjit Soni,
{\it Probing flavor changing top-charm-scalar interactions in $e^+e^-$ collisions},
Phys. Rev. D 53 (1996) 1199-1201, [arXiv:hep-ph/9506243].

\bibitem{tqv_ILC2}
Tim Tait, C.-P. Yuan, Phys. Rev. D 55 (1997) 7300; {\it Anomalous t-c-g coupling:
The connection between single top quark production and top quark decay},
Phys. Rev. D 55 (1997) 73007301.

\bibitem{tqv_ILC3}
V.F. Obraztsov, S.R. Slabospitsky, O.P. Yushchenko,
{\it Search for Anomalous Top-Quark Interaction at LEP-2 Collider},
Phys. Lett. B 426 (1998) 393-402, [arXiv:hep-ph/9712394].

\bibitem{tqv_ILC4}
Tao Han, JoAnne L. Hewett, {\it Top-Charm Associated Production in High Energy
$e^+e^-$ Collisions}, Phys. Rev. D 60 (1999) 074015, [arXiv:hep-ph/9811237].

\bibitem{tqv_ILC5}
Jiang Yi, Zhou Mian-Lai, Ma Wen-Gan, Han Liang, Zhou Hong, Han Meng,
{\it Probing flavor-changing interactions in photon-photon collisions},
Phys. Rev. D 57 (1998) 43434351, [arXiv:hep-ph/0003175].

\bibitem{tqv_ILC6}
Junjie Cao, Zhaohua Xiong, Jin Min Yang, {\it SUSY-Induced Top Quark FCNC
Processes at Linear Colliders}, Nucl. Phys. B 651 (2003) 87-105, [arXiv:hep-ph/0208035];
Junjie Cao, Guoli Liu, Jin Min Yang,
{\it Probing New Physics from Top-charm Associated Productions at Linear Colliders},
Eur. Phys. J. C 41 (2005) 381-391, [arXiv:hep-ph/0311166].

\bibitem{tqv_hadronee}
J. A. Aguilar-Saavedra, {\it Top flavour-changing neutral interactions:
theoretical expectations and experimental detection},
Acta Phys. Polon. B 35 (2004) 2695-2710, [arXiv:hep-ph/0409342];
J. A. Aguilar-Saavedra, {\it Top flavour-changing neutral coupling signals at a linear collider},
Phys. Lett. B 502 (2001) 115-124, [ arXiv:hep-ph/0012305 ].

\bibitem{tqv_ep1}
Orhan Cakir, {\it Anomalous production of top quarks at CLIC+LHC based gamma p colliders},
J. Phys. G 29 (2003) 1181-1192.

\bibitem{tqv_ep2}
A. T. Alan, A. Senol, {\it Single Top Production at HERA and THERA},
Europhys. Lett. 59 (2002) 669-673, [arXiv:hep-ph/0202119].

\bibitem{tqv_ep3}
A. A. Ashimova, S. R. Slabospitsky,
{\it The Constraint on FCNC Coupling of the Top Quark with a Gluon from ep Collisions},
Phys. Lett. B 668 (2008) 282-285, [arXiv:hep-ph/0604119].

\bibitem{tqv_ep4}
F. D. Aaron et cl., (H1 Collaboration), {\it Search for Single Top Quark Production at HERA},
Phys. Lett. B 678 (2009) 450-458, [arXiv:0904.3876].

\bibitem{tqv_pp1}
Ehab Malkawi, Tim Tait, {\it Top-quark-charm-quark strong flavor-changing neutral
currents at the Fermilab Tevatron}, Phys. Rev. D 54 (1996) 57585762.

\bibitem{tqv_pp2}
F. del Aguila, J. A. Aguilar-Saavedra, Ll. Ametller,
{\it Zt and $\gamma$t production via top flavour-changing neutral couplings at the Fermilab Tevatron},
Phys. Lett. B 462 (1999) 310-318, [arXiv:hep-ph/9906462].
F. del Aguila, J. A. Aguilar-Saavedra,
{\it Multilepton production via top flavour-changing neutral couplings at the CERN LHC},
Nucl. Phys. B 576 (2000) 56-84, [arXiv:hep-ph/9909222].

\bibitem{tqv_pp3}
Jaume Guasch, Wolfgang Hollik, Siannah Penaranda, Joan Sola,
{\it Single top-quark production by direct supersymmetric flavor-changing
neutral-current interactions at the LHC},
Nucl. Phys. Proc. Suppl. 157 (2006) 152-156, [arXiv:hep-ph/0601218].

\bibitem{tqv_pp4}
T. Han, R. D. Peccei, X. Zhang, {\it Top quark decay via flavor changing neutral currents at hadron colliders },
Nucl. Phys. B 454, 527 (1995);
T. Han, M. Hosch, K. Whisnant, Bing-Lin Young, X. Zhang,
{\it Single Top Quark Production via FCNC Couplings at Hadron Colliders},
Phys.Rev. D 58 (1998) 073008, [arXiv:hep-ph/9806486].

\bibitem{tqv_pp5}
P. M. Ferreira, R. B. Guedes, R. Santos, {\it Combined effects of strong and
electroweak FCNC effective operators in top quark physics at the LHC},
Phys. Rev. D 77 (2008) 114008, [ arXiv:0802.2075].

\bibitem{tqv_pp6}
Junjie Cao, Zhaoxia Heng, Lei Wu, Jin Min Yang,
{\it R-parity violating effects in top quark flavor-changing neutral-current production at LHC},
Phys. Rev. D 79 (2009) 054003, [ arXiv:0812.1698];
J.J.Cao, G.Eilam, M.Frank, K.Hikasa, G.L.Liu, I.Turan, J. M. Yang,
{\it SUSY-induced FCNC top-quark processes at the Large Hadron Collider},
Phys. Rev. D 75 (2007) 075021, [arXiv:hep-ph/0702264];
Junjie Cao, Gad Eilam, Ken-ichi Hikasa, Jin Min Yang,
{\it Experimental Constraints on Scharm-Stop Flavor Mixing and Implications in Top-quark FCNC Processes},
Phys. Rev. D 74 (2006) 031701, [arXiv:hep-ph/0604163];
Xiao-Fang Han, Lei Wang, Jin Min Yang,
{\it Top quark FCNC decays and productions at LHC in littlest Higgs model with T-parity},
Phys. Rev. D 80 (2009) 015018, [arXiv:0903.5491].

\bibitem{tqv_pp7}
M. Hosch, K. Whisnant, and B.-L. Young, {\it Direct top quark production
at hadron colliders as a probe of new physics}, Phys. Rev. D56, 5725 (1997).

\bibitem{tqv_pp8}
Nikolaos Kidonakis, Ben D. Pecjak, {\it Top-quark production and QCD},
Eur. Phys. J. C 72 (2012) 2084, [arXiv:1108.6063];
Nikolaos Kidonakis, Elwin Martin,
{\it FCNC Top Quark Production via Anomalous Couplings}, [arXiv:1310.0363].

\bibitem{tqv_pp9}
Jaume Guasch, Wolfgang Hollik, Siannah Penaranda, Joan Sola,
{\it Single top-quark production by direct supersymmetric flavor-changing neutral-current interactions at the LHC},
Nucl. Phys. Proc. Suppl. 157 (2006) 152-156, [arXiv:hep-ph/0601218].

\bibitem{tqv_pp10}
David Lopez-Val, Jaume Guasch, Joan Sola,
{\it Single top-quark production by strong and electroweak supersymmetric flavor-changing interactions at the LHC},
JHEP 0712 (2007) 054, [arXiv:0710.0587].

\bibitem{tqv_ppNLO1}
Nikolaos Kidonakis, Alexander Belyaev,
{\it FCNC top quark production via anomalous tqV couplings beyond leading order},
JHEP 0312 (2003) 004, [arXiv:hep-ph/0310299].

\bibitem{tqv_ppNLO2}
Yue Zhang, Bo Hua Li, Chong Sheng Li, Jun Gao, Hua Xing Zhu,
{\it Next-to-leading order QCD corrections to the top quark associated with $¦Ã$ production via model-independent flavor-changing neutral-current couplings at hadron colliders},
Phys. Rev. D 83 (2011) 094003, [arXiv:1101.5346];
Jun Gao, Chong Sheng Li, Jia Jun Zhang, Hua Xing Zhu,
{\it Next-to-leading order QCD corrections to the single top quark production via model-independent t-q-g flavor-changing neutral-current couplings at hadron colliders},
Phys. Rev. D 80 (2009) 114017, [arXiv:arXiv:0910.4349].

\bibitem{Anomaloustqr}
M.K\"{o}ksal, S. C. Inan, {\it Anomalous $tq\gamma$ couplings in
$\gamma p$ collision at the LHC}, [arXiv:1305.7096].

\bibitem{rbWtVtb1}
S. Ovyn, J. de Favereau de Jeneret,
{\it High energy single top photoproduction at the LHC}, [arXiv:0806.4841].

\bibitem{rbWtVtb2}
J. de Favereau de Jeneret, S. Ovyn,
{\it Single top quark photoproduction at the LHC},
Nucl. Phys. Proc. Suppl. 179-180 (2008) 277-284, [ arXiv:0806.4886].

\bibitem{AnomalousWtb}
B. Sahin, A. A. Billur, {\it Anomalous Wtb couplings in $\gamma$-proton
collision at the LHC}, Phys. Rev. D 86 (2012) 074026, [arXiv:1210.3235].

\bibitem{ppllpp1}
A. Abulencia et al., (CDF Collaboration),
{\it Observation of Exclusive Electron-Positron Production in Hadron-Hadron Collisions},
Phys. Rev. Lett. 98 (2007) 112001, [arXiv:hep-ex/0611040].

\bibitem{ppllpp2}
T. Aaltonen et al., (CDF Collaboration),
{\it Search for Exclusive Z-Boson Production and Observation of High-Mass
$p\bar{p}\rightarrow p\gamma\gamma\bar{p}\rightarrow pl^+l^- \bar{p}$
Events in $p\bar{p}$ Collisions at $\sqrt{s}=1.96$ TeV},
Phys. Rev. Lett. 102 (2009) 222002, [arXiv:0902.2816].

\bibitem{pprrpp}
T. Aaltonen et al., (CDF Collaboration),
{\it Search for Exclusive $\gamma\gamma$ Production in Hadron-Hadron Collisions},
Phys. Rev. Lett. 99 (2007) 242002, [arXiv:0707.2374].

\bibitem{ppjjpp}
T. Aaltonen et al., (CDF Collaboration),
{\it Observation of exclusive dijet production at the Fermilab Tevatron $\bar{p}p$ collider},
Phys. Rev. D. 77 (2008) 052004, [arXiv:0712.0604].

\bibitem{ppJPHIpp}
T. Aaltonen et al., (CDF Collaboration),
{\it Observation of Exclusive Charmonium Production and
$\gamma\gamma\rightarrow \mu^+\mu^-$ in $p\bar{p}$ Collisions at $\sqrt{s}=1.96$ TeV},
Phys. Rev. Lett. 102 (2009) 242001, [arXiv:0902.1271].

\bibitem{FDs1}
M. Tasevsky, {\it Diffractive physics program in ATLAS experiment},
Nucl. Phys. Proc. Suppl. 179-180 (2008) 187-195;
{\it Measuring Central Exclusive Processes at LHC}, [arXiv:0910.5205].

\bibitem{FDs2}
C. Royon, (RP220 Collaboration), {\it Project to install
roman pot detectors at 220 m in ATLAS}, [arXiv:0706.1796];
M.G. Albrow et al., {\it FP420: An $\&$ proposal to investigate
the feasibility of installing proton tagging detectors
in the 420-m region at LHC}, CERN-LHCC-2005-025 (2005);
B.E. Cox, (FP420 R and D Collaboration),
{\it The FP420 R$\&$D Project at the LHC}, [arXiv:hep-ph/0609209].

\bibitem{AFP}
C. Royon, {\it The ATLAS Forward Physics Project}, [arXiv:1302.0623].

\bibitem{HEPhotonIntatLHC}
J. de Favereau de Jeneret, V. Lemaitre, Y. Liu, S. Ovyn, T. Pierzchala,
K. Piotrzkowski, X. Rouby, N. Schul, M. Vander Donckt,
{\it High energy photon interactions at the LHC}, [arXiv:0908.2020].

\bibitem{SMWH}
S. Ovyn,{\it Associated W and Higgs boson photoproduction
and other electroweak photon induced processes at the LHC},
Nucl. Phys. Proc. Suppl. 179-180 (2008) 269-276, [arXiv:0806.1157].

\bibitem{SUSYprrp1}
N. Schul, K. Piotrzkowski, {\it Detection of two-photon
exclusive production of supersymmetric pairs at the LHC},
Nucl. Phys. Proc. Suppl. 179-180 (2008) 289-297, [arXiv:0806.1097];
K. Piotrzkowski, N. Schul,
{\it Two-photon exclusive production of supersymmetric pairs at the LHC},
AIP Conf.Proc. 1200 (2010) 434-437, [arXiv:0910.0202].

\bibitem{SUSYprrp2}
S. Heinemeyer, V. A. Khoze, M. G. Ryskin, W. J. Stirling, M. Tasevsky, G. Weiglein,
{\it Studying the MSSM Higgs sector by forward proton tagging at the LHC},
Eur. Phys. J. C 53 (2008) 231-256, [arXiv:0708.3052];
{\it Central Exclusive Diffractive MSSM Higgs-Boson Production at the LHC},
J. Phys. Conf. Ser. 110 (2008) 072016, [arXiv:0801.1974];
S. Heinemeyer, V.A. Khoze, M.G. Ryskin, M. Tasevsky, G. Weiglein,
{\it BSM Higgs Physics in the Exclusive Forward Proton Mode at the LHC},
Eur. Phys. J. C 71 (2011) 1649, [arXiv:1012.5007];
Marek Tasevsky, {\it Exclusive MSSM Higgs production at the LHC after Run I},
Eur. Phys. J. C 73 (2013) 2672, [arXiv:1309.7772].

\bibitem{EDpllp1}
S. Atag, S.C. Inan, I. Sahin,
{\it Extra dimensions in photon-induced two lepton final states at the CERN-LHC},
Phys. Rev. D 80 (2009) 075009, [arXiv:0904.2687].

\bibitem{EDprrp2}
S. Atag, S.C. Inan, I. Sahin, {\it Extra dimensions in
$\gamma\gamma\rightarrow\gamma\gamma$ process at the CERN-LHC},
JHEP 09 (2010) 042, [arXiv:1005.4792].

\bibitem{EDrqrq3}
I. Sahin, A.A. Billur, S.C. Inan, B. Sahin, M. Koksal, P. Tektas, E. Alici, R. Yildirim,
{\it Probe of extra dimensions in $\gamma q \rightarrow \gamma q$ at the LHC },
Phys. Rev. D 88 (2013) 095016, [arXiv:1304.5737].

\bibitem{unparticle}
I. Sahin, S. C. Inan, {\it Probe of unparticles at the LHC
in exclusive two lepton and two photon production via photon-photon fusion},
JHEP 0909 (2009) 069, [arXiv:0907.3290].

\bibitem{TTMrbtp}
Hao Sun, Chong-Xing Yue, {\it Precise photoproduction of the charged top-pions
at the LHC with forward detector acceptances}, Eur. Phys. J. C 74 (2014) 2823, [arXiv:1401.0250].

\bibitem{anoWWr1}
T. Pierzcha la, K. Piotrzkowski, {\it Sensitivity to
anomalous quartic gauge couplings in photon-photon interactions at the LHC},
Nucl.Phys.Proc.Suppl. 179-180 (2008) 257-264, [arXiv:0807.1121].

\bibitem{anoWWr2}
O. Kepka, C. Royon, {\it Anomalous $WW\gamma$ coupling
in photon-induced processes using forward detectors at the CERN LHC},
Phys. Rev. D 78 (2008) 073005, [arXiv:0808.0322].

\bibitem{anoVVV}
C. Royon, E. Chapon, O. Kepka,
{\it Anomalous trilinear and quartic WW$\gamma$, WW$\gamma\gamma$,
ZZ$\gamma$ and ZZ$\gamma\gamma$ couplings in photon induced processes at the LHC},
PoS EPS-HEP2009 (2009) 380, [arXiv:0909.5237].

\bibitem{anoWWrr}
E. Chapon, C. Royon, O. Kepka,
{\it Anomalous quartic $WW\gamma\gamma$, $ZZ\gamma\gamma$, and trilinear
$WW\gamma$ couplings in two-photon processes at high luminosity at the LHC},
Phys. Rev. D 81 (2010) 074003, [arXiv:0912.5161].

\bibitem{anoWWr3}
I. Sahin, A. A. Billur, {\it Anomalous $WW\gamma$ couplings in $\gamma-proton$
collision at the LHC}, Phys. Rev. D 83 (2011) 035011, [arXiv:1101.4998].

\bibitem{anoZZZ}
Rick S. Gupta, {\it Probing quartic neutral gauge boson couplings
using diffractive photon fusion at the LHC},
Phys. Rev. D 85 (2012) 014006, [arXiv:1111.3354].

\bibitem{anoZZrr}
I. Sahin, B. Sahin, {\it Anomalous quartic $ZZ\gamma\gamma$ couplings
in $\gamma proton$ collision at the LHC},
Phys. Rev. D 86 (2012) 115001, [arXiv:1211.3100].

\bibitem{anoZZrZrr}
A. Senol, {\it $ZZ\gamma$ and $Z\gamma\gamma$ anomalous couplings
in $\gamma p$ collision at the LHC},
Phys. Rev. D 87 (2013) 073003, [arXiv:1301.6914].

\bibitem{anoWWrrZZrrrp}
A. Senol, {\it Anomalous quartic $WW\gamma\gamma$ and $ZZ\gamma\gamma$
couplings in $\gamma p$ collision at the LHC}, [arXiv:1311.1370].

\bibitem{anoVVVV}
S. Fichet, G. von Gersdorff, O. Kepka, B. Lenzi, C. Royon, M. Saimpert,
{\it Probing new physics in diphoton production with proton tagging
at the Large Hadron Collider}, [arXiv:1312.5153].

\bibitem{electromagnetic1}
I. Sahin, M. Koksal, {\it Search for electromagnetic properties of
the neutrinos at the LHC}, JHEP 1103 (2011) 100, [arXiv:1010.3434].

\bibitem{electromagnetic2}
S. Atag, A.A. Billur, {\it Possibility of Determining $\tau$
Lepton Electromagnetic Moments in $\gamma\gamma\rightarrow \tau^{+}\tau^{-}$
Process at the CERN-LHC}, JHEP 1011 (2010) 060, [arXiv:1005.2841].

\bibitem{electromagnetic3}
I. Sahin, {\it Electromagnetic properties of the neutrinos in $\gamma$-proton
collision at the LHC}, Phys. Rev. D 85 (2012) 033002, [arXiv:1201.4364].

\bibitem{TripletH}
M. Chaichian, P. Hoyer, K. Huitu, V.A. Khoze, A.D. Pilkington,
{\it Searching for the triplet Higgs sector via central exclusive production at the LHC},
JHEP 0905 (2009) 011, [arXiv:0901.3746].

\bibitem{tqr_NLO}
J. J. Zhang, C. S. Li, J. Gao, H. Zhang, Z. Li, C. -P. Yuan, T. -C. Yuan,
{\it Next-to-leading order QCD corrections to the top quark decay via model-independent FCNC couplings},
Phys. Rev. Lett. 102 (2009) 072001, [arXiv:0810.3889];
{\it Next-to-leading order QCD corrections to the top quark decay
via the Flavor-Changing Neutral-Current operators with mixing effects},
Phys. Rev. D82 (2010) 073005, [arXiv:1004.0898];
Y. Zhang, B. H. Li, C. S. Li, J. Gao, H. X. Zhu,
{\it Next-to-leading order QCD corrections to the top quark associated
with $\gamma$ production via model-independent flavor-changing neutral-current
couplings at hadron colliders}, Phys. Rev. D 83 (2011) 094003, [arXiv:1101.5346].
J. Drobnak, S. Fajfer, J. F. Kamenik, {\it Flavor Changing Neutral Coupling Mediated Radiative
Top Quark Decays at Next-to-Leading Order in QCD},
Phys. Rev. Lett. 104 (2010) 252001, [arXiv:1004.0620];
{\it QCD Corrections to Flavor Changing Neutral Coupling Mediated Rare Top Quark Decays},
Phys. Rev. D 82 (2010) 073016. [arXiv:1007.2551].

\bibitem{twb_NLO}
C. S. Li, R. J. Oakes, T. C. Yuan, {\it QCD corrections to $t\rightarrow W^+b$ },
Phys. Rev. D 43 (1991) 3759-3762.

\bibitem{tqr_CDF}
F. Abe et al., (CDF Collaboration),
{\it Search for Flavor-Changing Neutral Current Decays of the Top Quark
in $p\bar{p}$ Collisions at $\sqrt{s}$ = 1.8 TeV},
Phys. Rev. Lett. 80 (1998) 2525-2530.

\bibitem{tqr_ZEUS}
S. Chekanov et al., (ZEUS Collaboration),
{\it Search for single-top production in ep collisions at HERA},
Phys. Lett. B 559 (2003) 153-170, [arXiv:hep-ex/0302010].

\bibitem{tqr_H1}
S. Dusini et al., (H1 and ZEUS Collaborations),
{\it Single top production and FCNC processes},
Nucl. Phys. Proc. Suppl. 109 B (2002) 262-265.

\bibitem{anomalous_ATLAS}
J. Carvalho et al. (ATLAS Collaboration),
Eur. Phys. J. C 52 (2007) 999-1019, [arXiv:0712.1127];
Veloso Filipe Manuel Almeida, Carvalho Jo$\tilde{a}$o, Onofre Ant$\acute{o}$nio,
{\it Study of ATLAS sensitivity to FCNC top quark decays}, CERN-THESIS-2008-106.

\bibitem{anomalous_CMS}
 L. Benucci, A. Kyriakis, {\it CMS sensitivity to top flavour changing neutral currents},
 Nucl. Phys. Proc. Suppl. 177-178 (2008) 258-260.

\bibitem{anomalous_LHC}
Efe Yazgan et al., (ATLAS Collaboration, CDF Collaboration, CMS Collaboration, D0 Collaboration),
{\it Flavor changing neutral currents in top quark production and decay}, CMS CR-2013/398, [arXiv:1312.5435].

\bibitem{EPA}
V. M. Budnev, I. F. Ginzburg, G. V. Meledin, V. G. Serbo,
{\it The two-photon particle production mechanism. Physical problems.
Applications. Equivalent photon approximation},
Phys. Rep. 15 (1975) 181;
G. Baur, K. Hencken, D. Trautmann, S. Sadowsky and Y. Kharlov,
{\it Coherent $\gamma\gamma$ and $\gamma A$ interactions in very peripheral
collisions at relativistic ion colliders},
Phys. Rep. 364 (2002) 359, [arXiv:hep-ph/0112211];
K. Piotrzkowski, {\it Tagging two-photon production at the CERN Large Hadron Collider},
Phys. Rev. D 63 (2001) 071502, [hep-ex/0009065].

\bibitem{FeynArts}
T. Hahn, {\it Generating Feynman diagrams and amplitudes with FeynArts 3},
Comput. Phys. Commun. 140 (2001) 418-431, [arXiv:hep-ph/0012260].

\bibitem{FormCalc}
T. Hahn, {\it Automatic Loop Calculations with FeynArts, FormCalc, and LoopTools},
Nucl. Phys. Proc. Suppl. 89 (2000) 231-236, [arXiv:hep-ph/0005029];
S. Agrawal, T. Hahn, E. Mirabella, {\it FormCalc 7},
J. Phys. Conf. Ser. 368 (2012) 012054, [arXiv:1112.0124].

\bibitem{LoopTools}
T. Hahn, M. Perez-Victoria, {\it Automatized one loop calculations in four-dimensions and D-dimensions},
Comput. Phys. Commun. 118 (1999) 153-165, [arXiv:hep-ph/9807565].

\bibitem{CT10}
Marco Guzzi, Pavel Nadolsky, Edmond Berger, Hung-Liang Lai, Fredrick Olness, C.-P. Yuan,
{\it CT10 parton distributions and other developments in the global QCD analysis},
SMU-HEP-10-11, [arXiv:1101.0561].

\bibitem{BASES}
S. Kawabata, {\it A new version of the multi-dimensional integration
and event generation package BASES/SPRING}, Comp. Phys. Commun 88 (1995) 309-326;
F. Yuasa, D. Perret-Gallix, S. Kawabata, T. Ishikawa, {\it Pvm-grace},
Nucl. Instrum. Meth. A 389 (1997) 77-80.

\bibitem{Kaleu}
Hameren A van, {\it Kaleu: a general-purpose parton-level phase space generator},
[arXiv:1003.4953].

\bibitem{2012PDG}
J. Beringer et al., (Particle Data Group),
{\it Review of Particle Physics (RPP)}, Phys. Rev. D 86 (2012) 010001.

\bibitem{xi123}
M. G. Albrow et al., (FP420 R and D Collaboration),
{\it The FP420 R$\&$D Project: Higgs and New Physics with forward protons at the LHC},
[arXiv:0806.0302].

\bibitem{misjets}
S. Chatrchyan et al., (CMS Collaboration),
{\it Identification of b-quark jets with the CMS experiment},
JINST 8 (2013) P04013, [arXiv:1211.4462].

\bibitem{SSformula}
G. L. Bayatian et al., (CMS Collaboration), J. Phys. G 34 (2007) 995.

\bibitem{tqr_limit1}
Xin-Qiang Li, Ya-Dong Yang, Xing-Bo Yuan,
{\it Anomalous $tq\gamma$ coupling effects in exclusive radiative B-meson decays},
JHEP 1108 (2011) 075, [arXiv:1105.0364].

\bibitem{tqr_limit2}
Xing-Bo Yuan, Yang Hao, Ya-Dong Yang,
{\it $\bar{B} \rightarrow X_s \gamma$ constraints on the top
quark anomalous $t\rightarrow c\gamma$ coupling},
Phys. Rev. D 83 (2011) 013004, [arXiv:1010.1912].


\end{thebibliography}
\end{document}